\documentclass[10pt,final,journal,twocolumn]{IEEEtran}
\usepackage[lined,boxruled]{algorithm2e}
\usepackage{algorithmic}
\usepackage{pifont}% http://ctan.org/pkg/pifont
\newcommand{\cmark}{\ding{51}}%
\newcommand{\xmark}{\ding{55}}%
\usepackage{float}
\usepackage{xspace}
\usepackage{textcomp}
\usepackage{fancyhdr}
\usepackage{amssymb,amsmath, amsfonts}
\usepackage{multirow}
\usepackage{textcomp} 
\usepackage{bbding}
\usepackage{color}
\usepackage{pifont}
\usepackage{comment}
\usepackage[hyphens]{url}
\usepackage{cite}
\usepackage{graphicx}
\usepackage{enumitem}
\usepackage{colortbl}
\newcommand{\fakeparagraph}[1]{\vspace{.1mm}\noindent\textbf{#1}}
\newcommand{\fakepar}[1]{\fakeparagraph{#1}}
\begin{document}
\font\myfont=cmr12 at 20pt

\title{\myfont Industrial Digital Twins at the Nexus of NextG Wireless Networks and Computational Intelligence: A Survey}

\author{Shah~Zeb,~Aamir~Mahmood,~Syed~Ali~Hassan,~MD.~Jalil~Piran,\\
	Mikael~Gidlund,~and~Mohsen~Guizani.% <-this % stops a space
	\thanks{S. Zeb and S. A. Hassan are with the School of Electrical Engineering and Computer Science (SEECS), National University of Science and Technology (NUST), Pakistan, (email: \{szeb.dphd19seecs, ali.hassan\}@seecs.edu.pk).
    
    A. Mahmood and M. Gidlund are with the Department of Information Systems \& Technology, Mid Sweden University, Sweden, (email: \{firstname.lastname\}@miun.se).
    
    MD. J. Piran is with the Department of Computer Science and Engineering, Sejong University, Seoul, South Korea, (email: piran@sejong.ac.kr).
    
    M. Guizani is with the Department of Computer Science and Engineering, Qatar University, Doha, Qatar, (email: mguizani@ieee.org).
    }
	\vspace{-20pt}
	}

\maketitle

\begin{abstract}
By amalgamating recent communication and control technologies, computing and data analytics techniques, and modular manufacturing, Industry~4.0 promotes integrating cyber-physical worlds through cyber-physical systems (CPS) and digital twin (DT) for monitoring, optimization, and prognostics of industrial processes.
A DT is an emerging but conceptually different construct than CPS. Like CPS, DT relies on communication to create a highly-consistent, synchronized digital mirror image of the objects or physical processes. DT, in addition, uses built-in models on this precise image to simulate, analyze, predict, and optimize their real-time operation using feedback. DT is rapidly diffusing in the industries with recent advances in the industrial Internet of things (IIoT), edge and cloud computing, machine learning, artificial intelligence, and advanced data analytics. However, the existing literature lacks in identifying and discussing the role and requirements of these technologies in DT-enabled industries from the communication and computing perspective. 
In this article, we first present the functional aspects, appeal, and innovative use of DT in smart industries. Then, we elaborate on this perspective by systematically reviewing and reflecting on recent research in next-generation (NextG) wireless technologies (e.g., 5G and beyond networks), various tools (e.g., age of information, federated learning, data analytics), and other promising trends in networked computing (e.g., edge and cloud computing). Moreover, we discuss the DT deployment strategies at different industrial communication layers to meet the monitoring and control requirements of industrial applications. We also outline several key reflections and future research challenges and directions to facilitate industrial DT's adoption.
\end{abstract}
\vspace{-5pt}

\maketitle
\IEEEpeerreviewmaketitle
% https://medium.com/dataseries/digital-twin-solutions-c5288fcc3bc7
\section{Introduction}
\begin{figure}
\centering
\includegraphics[width=1\linewidth]{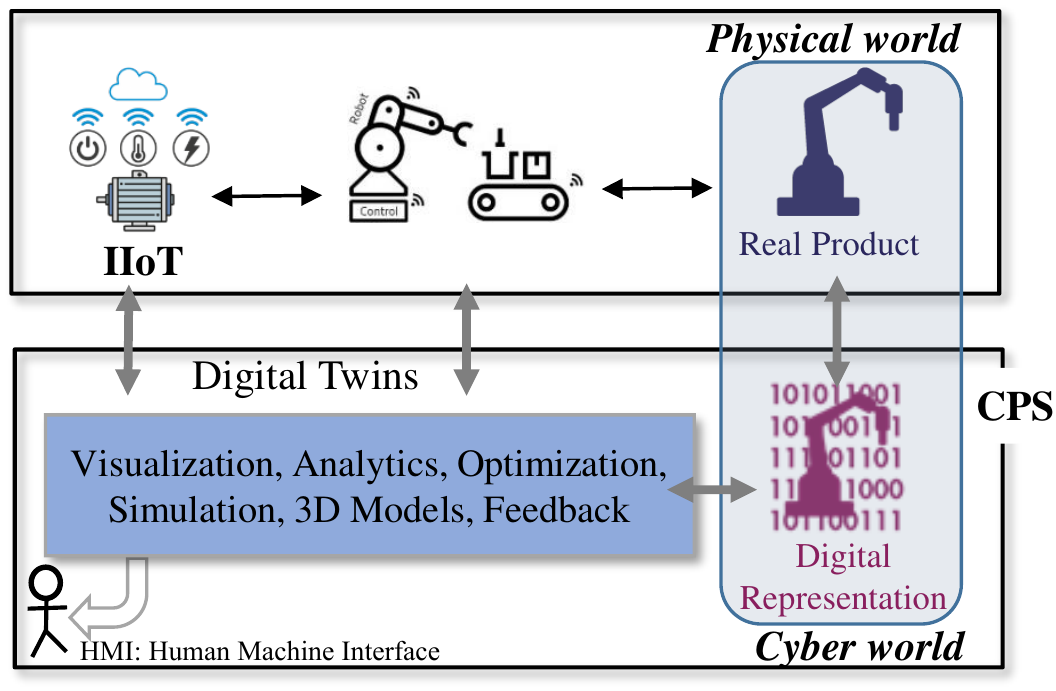}
	\caption{Illustration of IIoT, CPS and DT for real-world physical assets (e.g., sensors, machines, robots) and their digital representation, \textit{the simulated cyber space}.}
	\label{fig:DT}
 	\vspace{-10pt}
\end{figure}

\IEEEPARstart{T}he fourth industrial revolution, termed Industry 4.0, targets digital transformation of various sectors, such as intelligent manufacturing, automation, and aerospace~\cite{kourtis2019rule,random1}. In this transformation, the intelligent factory, also known as the factory of the future, depends on ubiquitous industrial Internet of things (IIoT) connectivity to achieve the goals of flexible, efficient, and versatile production systems. On the other hand, the emerging architectures, such as cyber-physical systems (CPS) and industrial digital twin (DT), together with the intelligent computation-enabled next-generation (NextG) wireless networks (i.e., 5G-and-Beyond networks), are envisioned to play a prominent role in reshaping the digital landscape of factories of the future. Fig.~\ref{fig:DT} illustrates the conceptual relation among IIoT, CPS, and DT for physical entities on a factory floor, which are further elaborated below.

\fakepar{Industrial IoT}: IoT is a revolutionary concept of building an intelligent digital ecosystem by connecting all physical assets empowered to interact or communicate through the Internet infrastructure and NextG wireless networks~\cite{piran2016qoe}. Meanwhile, integration of Industry 4.0 with IoT in the products' manufacturing process has given surge to IIoT---a child IoT technology designed explicitly for mission-critical industrial applications~\cite{munirathinam2020industry}. The connected industrial assets are machines, actuators, control systems, and robots, performing quality-of-service (QoS)-aware mission-oriented automation tasks. IIoT network differs from a typical ad-hoc IoT network; it is primarily data analytics-enabled cloud-based structured network infrastructure that supports machine-to-machine (M2M) wireless communications having stringent latency and reliability requirements in a dynamic industrial environment~\cite{sisinni2018industrial}. In industrial automation, the monitoring applications are typically not affected by the delay and jitter in packets, and the tolerable latency is in the order of seconds. However, critical processes such as closed-loop control and interlocking have stringent latency requirements of 1 ms to 100 ms and ultra-reliability of more than 99.999\%~\cite{mahmood2019time}. 

\fakepar{CPS and Industrial DT}: CPS brings together the physical and networked resources with emerging computation paradigm, enabling the intelligence in machines and robots to perform collaborative mission-critical tasks~\cite{zhang2018framework,random2}. Meanwhile, DT is a \textit{living} virtual or digital image/softwarized model that can be built for robots, machines, or the physical process of the entire manufacturing plant, which interacts with the physical assets of the plant using actuators and control planes to optimize the production~\cite{he2018surveillance,random5}. Essentially, it is a digital tool that recreates an intelligent virtual image of the machines in the edge or cloud based on the incoming IIoT data from field devices, associated with real-time physical attributes of a CPS. This implies that a DT can be implemented at various levels of the layered communication pyramid, i.e.,  at the edge close to the data sources or the cloud close to the application~\cite{leng2021digital}. In a nutshell, Industry 4.0 is the product of an amalgamation of two splendid paradigms, IIoT and the CPS, which is further aided by DT~\cite{random4,aazam2018deploying}. The headpin of this globally adopted industrial revolution is the unprecedented implementation of intelligent services using emerging technologies.

Currently, more industries are opting to adopt the DT-driven industrial paradigm thanks to the advancements in communication and sensing technologies, virtualization, and computing power in facilitating, customizing, and optimizing the factory processes and machines~\cite{wang2021new,stark2019development}. By 2025, more than six billion IoT-enabled devices will be online through cellular access, which currently stands at 1.5 billion connections, and the generated cellular traffic will reach $10^{18}$ bytes~\cite{ericsson2020, mmultimedia}. The COVID-19 pandemic has affected the high forecast of connections and online traffic predicted in previous technical reports~\cite{ericsson2020}. Nevertheless, little has changed as it has increased the demand for the acquisition of intelligent services that can be managed and controlled remotely through information and communication technologies (ICT)~\cite{6ga,chamola2020comprehensive,allam2021future}.
This strengthens the importance of emerging trending technologies and techniques in NextG wireless networks and computational intelligence paradigms as their nexus will provide the baseline for developing industrial DTs.

\fakepar{NextG Wireless Networks}:
The design and deployment of the fifth-generation (5G) and beyond (B5G) wireless networks is primarily focused on supporting diverse services with heterogeneous communication attributes of mission-critical applications~\cite{chettri2019comprehensive}. These communication attributes are~\cite{9430902,alahmad2021mobile,abbas2021age}: 1) ultra-reliability and low latency, 2) support for high data rates, 3) massive machine connectivity, 4) secure data-driven mobile computation services, 5) dynamic and optimized over-the-air resource allocations, and 6) energy-efficient green communication with minimum age-of-information (AoI). Collectively, these enabling attributes provided by the NextG wireless networks form the building foundation for two-way communication between industrial DT and the physical assets.

\fakepar{Computational Intelligence}:
The concurrent deployment of the enhanced networked communication infrastructure, high-performance data analytics (HPDA) techniques and high power computing (HPC) capabilities at the cloud/edge is ushering new \textit{computational intelligence} paradigms that can provide the customized services to on-demand industrial applications, e.g., anomalies detection, fault prognosis, and increased digital hyperconnectivity~\cite{tang2021computing,xiao2020toward}.
One of the new computational intelligence paradigms called "federated learning" combines the data analytics and computing models at the edge of network to provide intelligent services (data offloading, efficient computations) for the end-devices, e.g., IIoT-connected robots and machines~\cite{khan2021federated}. Similarly, data fusion and streaming analytics with HPC capabilities can provide real-time analysis of IIoT data.

\fakepar{Industrial DT at the Nexus}:
The nexus of NextG wireless networks and emerging computational intelligence paradigms in tandem is expected to play an essential role in realizing the true potential of Industrial DTs and bridging the cyber-space and physical space comprising multiple robots and machines. Many factory assets are expected to continuously transmit an ample amount of machine data to the HPC-enabled edge or cloud servers, utilizing NextG wireless networks' resources. Similarly, HPDA-based algorithms assist in realizing the softwarization of physical space based on the incurred IIoT data at the HPC-enabled edge or cloud servers to model the industrial DT. Once the industrial DT is modelled, it monitors, controls, and optimizes the industrial process with NextG wireless networks. This increases the significance of discussing the roles and requirements of the emerging computational and communication enablers in both NextG wireless networks and computational intelligence paradigms.

%%%%%%%%%%%%%%%%%%%%%%%%%%%%%%%%%%%%%%%%%%%%%%%%%
\section{Research Trends, Gaps in Existing Surveys, and Our Contributions }
This section discusses the market statistics and current research trends in critical enablers of Industry 4.0 (IIoT, CPS, and DT), research methodology for collecting and evaluating literature, summary of existing surveys and review works on digital twins in various industries, and motivation and contributions of our review work.
\subsection{Market Statistics and Research Trends}
\begin{figure}
\centering
\includegraphics[width=0.95\linewidth]{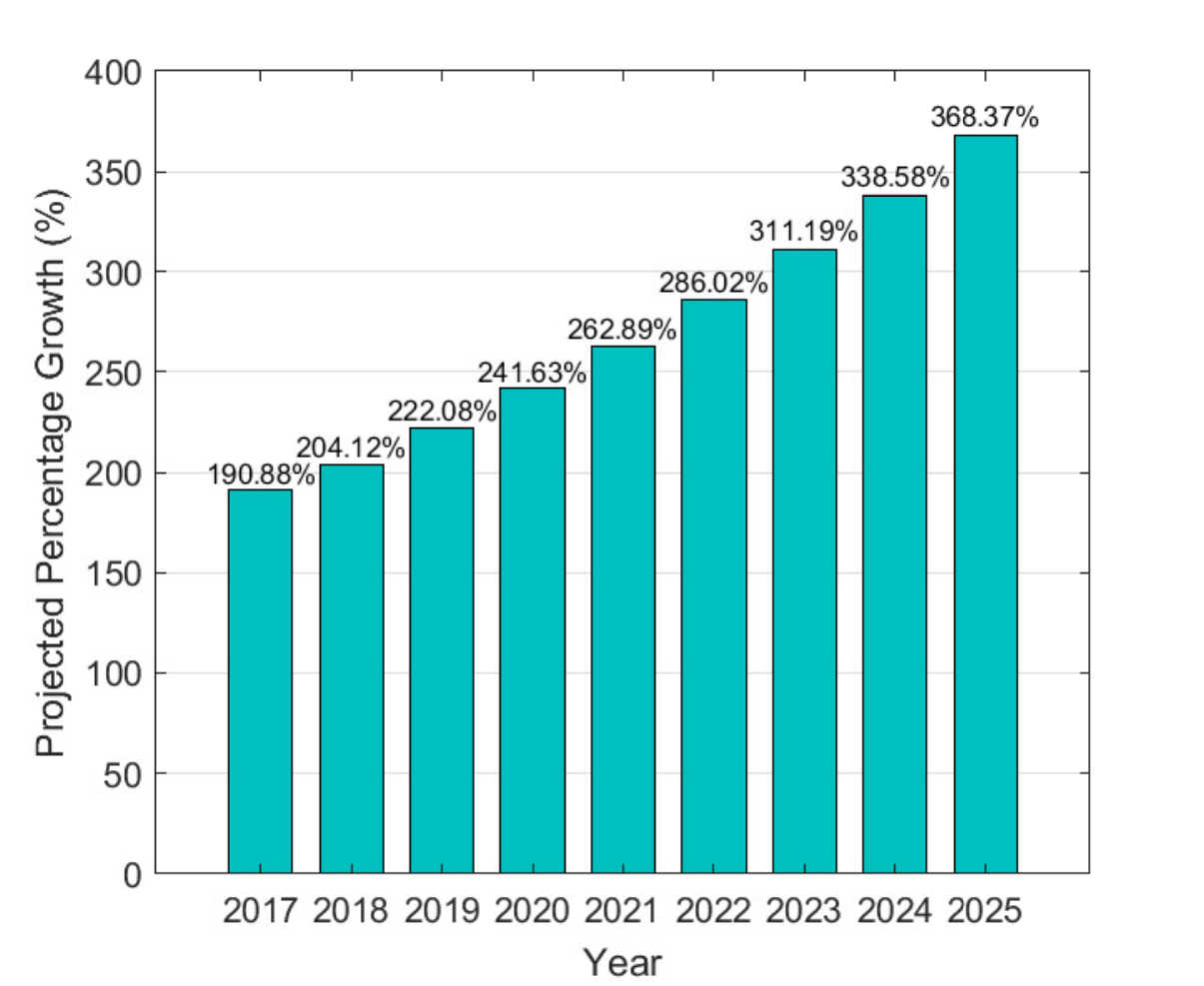}
	\caption{Annual expected size projection of the worldwide business by factory automation (2017-2025)~\cite{MG3}}
	\label{fig:market}
 	\vspace{-10pt}
\end{figure}

The trend to incorporate digitization and robotization in the manufacturing and aerospace sectors is growing rapidly to enhance agility and efficiency of the production processes~\cite{MG4}. This is apparent from the increasing density of robots on the factory floors; in developed countries, such as China, South Korea, and Germany, more than 500 industrial robots exist on average per 10000 employees. 
Meanwhile, the International Federation of Robotics (IFR) records
show that the worldwide number of operating robots is 2.7 million, showing an increase of 12\% from the previous year. In this emerging scenario, the factories of the future demand the networked interaction of collaborating multiple robots to perform isochronous and intelligent operations. The critical nature of these collaborative operations is becoming possible with the IIoT connectivity technologies together with the emerging CPS/DT-based synchronized digital breathing replicas~\cite{MG1,MG2}. These trends and technological advances have been drivers behind the global market increase in factory automation. According to the market statistics (c.f.~Fig.~\ref{fig:market}), the factory automation market is projected to grow exponentially at the compound annual growth rate of 8.8\% during the 2017-2025 time period with a forecasted value of 368 billion USD~\cite{MG3}.

\begin{figure}
\centering
\includegraphics[width=0.95\linewidth]{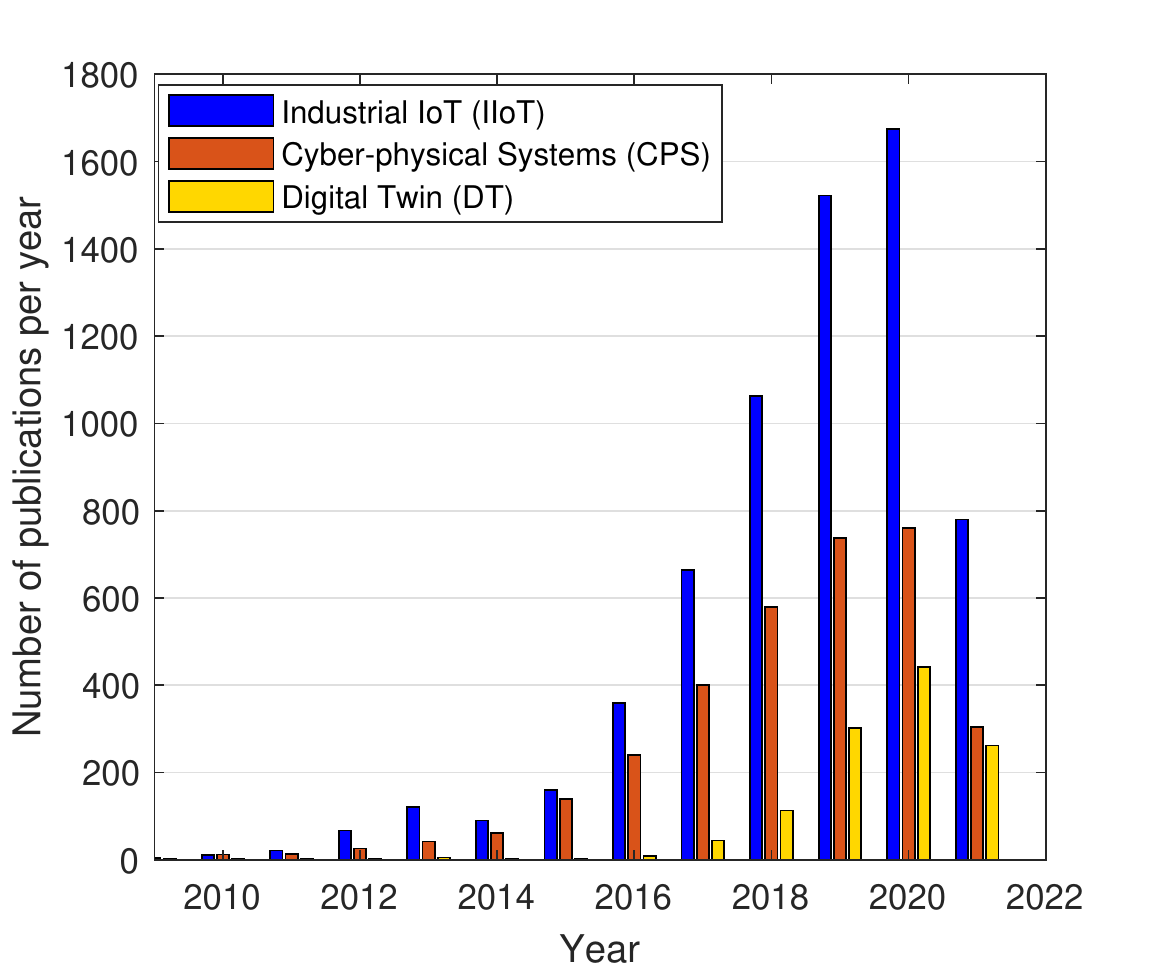}
	\caption{Year wise publications showing the research trends of IIoT, CPS, and DT [Scopus data, Access date: 10 July, 2021].}
	\label{fig:publications}
 	\vspace{-10pt}
\end{figure}

Meanwhile, it is apparent from Fig.~\ref{fig:publications} that significant growth has been observed in research publications every year for both CPS and DT during 2011-2021. 
{\renewcommand{\arraystretch}{1.1}
\begin{table*}[t!]
\centering
	\caption{Methodology on Screening Papers} 
\begin{tabular}{|c|c|}
\hline
 \textbf{Index of Searching} & \textbf{Content of Evaluation} \\ \hline
 Search Time-period & From: January 2003, To: July 2021 \\ \hline
 Article Database & Scopus \\ \hline
 Articles Type & Published peer-reviewed technical conferences and journals   \\ \hline
 Screening Procedures & \begin{tabular}[c]{@{}c@{}}The relevance with the research topic as judged by the content\\ written in the abstract, introduction and conclusion section of each paper.    \end{tabular}  \\ \hline
  Search Strings & \begin{tabular}[c]{@{}c@{}}"Industrial IoT", "IoT for Industry 4.0", "cyber-physical systems", \\ "digital twin", "digital twin manufacturing", digital twin and Industry 4.0", etc. \end{tabular}  \\ \hline
\end{tabular}
\label{methodology}
\vspace{-10pt}
\end{table*}
}
The screening methodology to  obtain Fig.~\ref{fig:publications} is summarized in Table.~\ref{methodology}. Note that we repeated the screening process through three different independent campaigns and compiled the findings from 2003-2021 for CPS, IIoT, and DT to bring reliability to the publication screening process. 
However, before final processing, the relevance of the compiled data with the area of interest needs to be established, i.e., it should be based on the abstract, introduction, and conclusion of the papers. Moreover, numerous articles in the search database included the keyword "digital" or "twin" in the abstract or title, which does not mean the "digital twin" or "virtual image" of the process as a whole. The same goes for "cyber" or "physical systems", and "industrial" or "IoT" while searching data for CPS and IIoT. Such types of articles were excluded from the final database used to plot the trend of publication count in Fig.~\ref{fig:publications}.

We read through all the incorporated papers in our search database and tried to find the common grounds and proposition towards CPS, IIoT, and DT in the industrial ecosystem.
During 2003-2011, there was significant adoption and development in IoT, sensor technology, machine analytics, simulation, and communication technologies, which provided a baseline for further work in the areas of CPS and IIoT.
However, the technological foundations were not mature enough to support DT deployment in industrial applications. Since 2011, there is a significant focus shift towards the DT research and development in tandem with IIoT and CPS, as evident from Fig.~\ref{fig:publications}. However, fewer attempts have been made to rigorously evaluate the DT application in the industry. Based on the facts mentioned above, DT is expected to open up novel opportunities for research and development in the foreseeable future.

\subsection{Existing Surveys and Review Works}
\definecolor{Gray}{gray}{0.9}
\definecolor{Gray1}{gray}{0.7}
\bgroup
{\renewcommand{\arraystretch}{1.1}
\begin{table*}[!ht]
\centering
\caption{Summary of existing surveys and case reviews on digital twin with their primary research focus. 
}
\scalebox{0.83}{
%\small\addtolength{\tabcolsep}{-5pt}
\begin{tabular}{|c|c|c|c|c|c|c|c|c|}
\hline
% \textbf{References} & \textbf{\begin{tabular}[c]{@{}c@{}}Smart Factory\\Application\end{tabular}} & \textbf{\begin{tabular}[c]{@{}c@{}}Cloud-Edge\\Computing\end{tabular}} & \textbf{\begin{tabular}[c]{@{}c@{}}Industrial\\ B5G Services\end{tabular}}  & \textbf{\begin{tabular}[c]{@{}c@{}}ML-\\ AI\end{tabular}} & \textbf{\begin{tabular}[c]{@{}c@{}}Big Data-\\ Fusion\end{tabular}} & \textbf{\begin{tabular}[c]{@{}c@{}}DT Placement \\ Strategies\end{tabular}} & \textbf{\begin{tabular}[c]{@{}c@{}}Green Comm-\\ unication\end{tabular}} & \textbf{\begin{tabular}[c]{@{}c@{}}AoI\end{tabular}} & \textbf{Remarks}  

% \multirow{3}{*}{\textbf{References}} & \multirow{3}{*}{\textbf{\begin{tabular}[c]{@{}c@{}}Smart Factory \\ Applications\end{tabular}}} & \multicolumn{3}{c|}{\textbf{Emerging Computation Enablers}} & \multicolumn{4}{c|}{\textbf{Emerging Communication Enablers}}  & \multirow{2}{*}{\textbf{Remarks}} \\ \cline{3-9}  &    & \textbf{\begin{tabular}[c]{@{}c@{}}Cloud-Edge\\Computing\end{tabular}} & \textbf{\begin{tabular}[c]{@{}c@{}}ML-\\ AI\end{tabular}} & \textbf{\begin{tabular}[c]{@{}c@{}}Big Data-\\ Fusion\end{tabular}} & \textbf{\begin{tabular}[c]{@{}c@{}}Industrial\\ B5G Services\end{tabular}} & \textbf{\begin{tabular}[c]{@{}c@{}}DT Placement \\ Strategies\end{tabular}} & \textbf{\begin{tabular}[c]{@{}c@{}}Green Comm-\\ unication\end{tabular}} & \textbf{\begin{tabular}[c]{@{}c@{}}AoI\end{tabular}} &                       
\multirow{3}{*}{\textbf{References}} & \multicolumn{3}{c|}{\textbf{Emerging Computation Enablers}}   & \multicolumn{4}{c|}{\textbf{Emerging Communication Enablers}}  & \multirow{3}{*}{\textbf{Remarks}} \\ \cline{2-8}                 & \textbf{\begin{tabular}[c]{@{}c@{}}Cloud-Edge\\ Computing\end{tabular}} & \textbf{\begin{tabular}[c]{@{}c@{}}ML-\\ AI\end{tabular}} & \textbf{\begin{tabular}[c]{@{}c@{}}Big Data-\\ Data Fusion\end{tabular}} & \textbf{\begin{tabular}[c]{@{}c@{}}Industrial\\ B5G Services\end{tabular}} & \textbf{\begin{tabular}[c]{@{}c@{}}DT Placement \\ Strategies\end{tabular}} & \textbf{\begin{tabular}[c]{@{}c@{}}Green Comm-\\ unication\end{tabular}} & \textbf{\begin{tabular}[c]{@{}c@{}}AoI\end{tabular}} &   

\\ \hline\hline

 \begin{tabular}[c]{@{}c@{}}Tao \textit{et al}., \\ \cite{tii1}\end{tabular}
&   \xmark  &  \xmark   &  \cmark & \xmark & \xmark  &   \xmark & \xmark   & \begin{tabular}[c]{@{}c@{}}Analyzed the latest DT review study to\\ better understand the development and \\implementation of DTs in industry.\end{tabular} \\ \hline

\rowcolor{Gray}
\begin{tabular}[c]{@{}c@{}}Qi \textit{et al}., \\ \cite{elsevier2}\end{tabular} 
&   \textbf{Cloud}  &  \cmark   & \cmark & \xmark & \xmark  &    \xmark & \xmark   & \begin{tabular}[c]{@{}c@{}}Provided the broad guidelines for \\DT enabling technology as well as \\particular tool examples.\end{tabular} \\ \hline
 
\begin{tabular}[c]{@{}c@{}}Cimino \textit{et al}., \\ \cite{elsevier3}\end{tabular} 
&  \textbf{Cloud}  &  \xmark   &    \cmark & \xmark & \xmark  &   \xmark & \xmark   & \begin{tabular}[c]{@{}c@{}} Investigated the uses of DT \\ in manufacturing and the accompanying \\ services provided by them.\end{tabular} \\ \hline
 
 \rowcolor{Gray}
 \begin{tabular}[c]{@{}c@{}}Yi \textit{et al}., \\ \cite{elsevier4}\end{tabular} 
&   \textbf{Cloud}  &  \xmark   &    \cmark & \xmark & \xmark  &   \xmark & \xmark  & \begin{tabular}[c]{@{}c@{}}Studied a DT reference model for\\ designing smart assembly processes.\end{tabular} \\ \hline
 
  \begin{tabular}[c]{@{}c@{}}Liu \textit{et al}., \\ \cite{elsevier5}\end{tabular} 
 &  \textbf{Cloud}  &  \cmark   & \cmark & \xmark & \xmark  &    \xmark & \xmark  & \begin{tabular}[c]{@{}c@{}}Reviewed and examined the previous\\ studies from the standpoint of DT ideas,\\ simulation technologies, and applications. \end{tabular} \\ \hline

\rowcolor{Gray}
\begin{tabular}[c]{@{}c@{}}Jones \textit{et al}., \\ \cite{elsevier7}\end{tabular} 
 &  \textbf{Cloud}  &  \xmark   &    \xmark & \xmark & \xmark  &   \xmark & \xmark  & \begin{tabular}[c]{@{}c@{}}Demonstrated a thorough review work on \\characterising the DT and its business value, \\highlighted research gaps and future prospects.\end{tabular} \\ \hline
 
 \begin{tabular}[c]{@{}c@{}}Tao \textit{et al}., \\ \cite{access12}\end{tabular} 
 &   \textbf{Cloud}  &  \cmark   &    \cmark & \xmark & \xmark  &   \xmark & \xmark  & \begin{tabular}[c]{@{}c@{}} Presented an overview of DT-based shop-floor \\ services as well as suggestions for future work.\end{tabular} \\ \hline
 
 \rowcolor{Gray}
 \begin{tabular}[c]{@{}c@{}}Qi \textit{et al}., \\ \cite{qi2018digital}\end{tabular} 
&  \textbf{Cloud}  &  \cmark   &    \cmark & \xmark & \xmark  &   \xmark & \xmark & \begin{tabular}[c]{@{}c@{}}Considered evaluating the roles of both\\ big data and DT, as well as their\\ interactions with smart manufacturing.\end{tabular} \\ \hline
 
 \begin{tabular}[c]{@{}c@{}}Rasheed \textit{et al}.,\\\cite{access14}\end{tabular} 
 &  \cmark  &  \cmark   &    \cmark & \xmark & \xmark  &   \xmark & \xmark  & \begin{tabular}[c]{@{}c@{}}Examined approaches and techniques relevant to\\ the creation of DT from a modelling viewpoint.\end{tabular} \\ \hline

\rowcolor{Gray}
 \begin{tabular}[c]{@{}c@{}}Fuller \textit{et al}.,\\\cite{access15}\end{tabular} 
  &   \textbf{Cloud}  &  \cmark   &    \cmark & \xmark & \xmark  &   \xmark & \xmark  & \begin{tabular}[c]{@{}c@{}}Surveyed the DT-related papers classified by\\ the type of research areas (smart cities, etc.).\end{tabular} \\ \hline
 
 \begin{tabular}[c]{@{}c@{}}Wanasinghe \textit{et}\\\textit{al}.,\cite{access16}\end{tabular} 
  &   \xmark  &  \cmark   &  \cmark & \xmark & \xmark  &   \xmark & \xmark  & \begin{tabular}[c]{@{}c@{}} Provided a literature overview of DT from the\\ standpoint of the Oil \& Gas industry, as well\\ as highlighted the future research objectives. \end{tabular} \\ \hline
 
 \rowcolor{Gray}
 \begin{tabular}[c]{@{}c@{}}Khajavi \textit{et al}.,\\\cite{access17}\end{tabular} 
 &  \textbf{Cloud}  &  \xmark   &  \textbf{Big Data} & \xmark & \xmark  &   \xmark & \xmark  & \begin{tabular}[c]{@{}c@{}}Considered DT for a building life cycle \\ management, and investigated the advantages \\and drawbacks.\end{tabular} \\ \hline
  
  \begin{tabular}[c]{@{}c@{}}Barricelli \textit{et al}.,\\\cite{access18}\end{tabular} 
 &  \xmark  &  \xmark   & \cmark  & \xmark & \xmark  &   \xmark & \xmark  & \begin{tabular}[c]{@{}c@{}}Reviewed current definitions, key features, and\\ socio-technical design elements in DT domains.\end{tabular} \\ \hline
 
 \rowcolor{Gray}
 \begin{tabular}[c]{@{}c@{}}Hasan \textit{et al}.,\\\cite{access19}\end{tabular} 
  &  \xmark  &  \xmark   & \xmark  & \xmark & \xmark  &   \xmark & \xmark & \begin{tabular}[c]{@{}c@{}} Discussed a blockchain-based DT creation\\ method to ensure the safe and reliable \\traceability of transactions, logs, etc. \end{tabular} \\ \hline
 
 \begin{tabular}[c]{@{}c@{}}Moyne \textit{et al}.,\\\cite{access20}\end{tabular} 
  &  \xmark  &  \xmark   & \cmark  & \xmark & \xmark  &   \xmark & \xmark   & \begin{tabular}[c]{@{}c@{}}Investigated requirements-based methodology for\\ determining baseline components for a framework\\ on which real DT solutions can be developed.\end{tabular} \\ \hline
 
 \rowcolor{Gray}
  \begin{tabular}[c]{@{}c@{}}Minerva \textit{et al}.,\\\cite{access21}\end{tabular} 
  &   \cmark  &  \xmark   & \cmark  & \xmark & \xmark  &   \xmark & \xmark   & \begin{tabular}[c]{@{}c@{}}Identified and reviewed a comprehensive DT\\ characteristics leading to the "virtualization" \\of physical space.\end{tabular} \\ \hline
 
 \begin{tabular}[c]{@{}c@{}}Rathore \textit{et al}.,\\\cite{access23}\end{tabular} 
  &   \cmark  &  \cmark   & \cmark  & \xmark & \xmark  &   \xmark & \xmark    & \begin{tabular}[c]{@{}c@{}}Conducted a thorough literature review on\\ DT systems that use ML \& AI technology.\end{tabular} \\ \hline
  
 \rowcolor{Gray}
 \begin{tabular}[c]{@{}c@{}}Zheng \textit{et al}.,\\\cite{zheng2019application}\end{tabular}
  &   \xmark  &  \xmark   & \textbf{Data Fusion}  & \xmark & \xmark  &   \xmark & \xmark  & \begin{tabular}[c]{@{}c@{}}Reviewed the related research, concept and app-\\ lication of DT technology, and proposed frame-\\ work of DT for product lifecycle management. \end{tabular} \\ \hline
 
 \begin{tabular}[c]{@{}c@{}}Wu \textit{et al}.,\\\cite{IOTJ25}\end{tabular}
  &  \cmark  &  \cmark   & \cmark  & \xmark & \xmark  &   \xmark & \xmark & \begin{tabular}[c]{@{}c@{}}Surveyed DT network to investigate the DT\\ significance in standard application scenarios.\end{tabular} \\ \hline

\rowcolor{Gray}
\textbf{Our Survey} &   \cmark   &  \cmark   & \cmark  & \cmark  & \cmark   &  \cmark  & \cmark    & \begin{tabular}[c]{@{}c@{}}Surveyed trending enabling technologies and \\techniques in communication and computation \\fields for industrial DT, and highlighted roles,\\requirements, and future research directions.\end{tabular} \\ \hline
 
\end{tabular}
}
\label{table:II}
% \vspace{-5pt}
\end{table*}
}
\egroup
Completeness is the priority of any review work. Numerous surveys and review works on various case studies primarily reviewed DT for control and management processes in industrial applications~\cite{tii1,elsevier2,elsevier3,elsevier4,elsevier5,elsevier7,access12,qi2018digital,access14,access15,access16,access17,access18,access19,access20,access21,access23,zheng2019application,IOTJ25}. The closely related works to this article are summarized in Table.~\ref{table:II} with the necessary emerging computation and communication enablers identified and marked for either they are covered in DT review work or not.
We use (\cmark) if the enabler technology is discussed and explored from the DT's factory usage perspective and (\xmark) otherwise.

The idea of survey work of authors in \cite{elsevier5,elsevier7,access14,access15} primarily centers around the: 1) DT concepts and characterization, 2) DT construction methodologies and modeling, 3) various applications of DT usage, 4) DT business value, and lastly, 5) provide the research gaps findings in DT literature and provide future research directions. Similarly, Cimino \textit{et al.} in \cite{elsevier3} explored the DT use cases in manufacturing sectors and identified the expected critical DT services on the factory management level.
Khajavi \textit{et al.} discusses the benefits and shortcomings of DT for building management, and developed the DT model for building using numerous IoT sensors and installed devices to manage the building life cycle~\cite{access17}. Furthermore, the authors of \cite{zheng2019application} reviewed the DT concepts and developed the DT-based management model for the product life cycle. Moyne \textit{et al.} in \cite{access20} identified the requirements of DT usage and developed a model based on the recommended requirements towards the practical implementation of DT. Minerva \textit{et al.} in \cite{access21} surveyed the DT features to enable softwarization and virtualization of physical objects and achieve true hyper-connectivity in application-specific environments, such as manufacturing industries.
Tao \textit{ et al.}, Qi \textit{et al.}, and the others in \cite{tii1,elsevier2,access23,qi2018digital} reviewed the DT usage for innovative factory applications. Key findings of their works are: 1) explore the DT research carried out for the industrial use cases, 2) identify critical DT enablers for implementation, i.e., cloud computation, big data, data fusion, ML, etc., and 3) interplay role of ML, AI, and big data is reviewed from the perspective of DT-based smart manufacturing floor. More review details are given in the remarks column of Table.~\ref{table:II}. Hasan~\textit{et al.} in \cite{access19} reviewed blockchain technology to implement the DT process and considered using a blockchain-based DT case study for securing the data transaction, logs, and other essential processes data. Likewise, Wanasinghe~\textit{et al.} in \cite{access16} performed a literature review for the use of DT technology in the oil and gas industry and explored the DT benefits, lapses, and identified future research directions.

From observing the review work in Table.~\ref{table:II}, most of the studies are only focused on exploring the computing enablers for DT. However, emerging technologies and techniques in communication and computing have not been explored together in the literature. Our review work focuses on industrial DT with respect to emerging state-of-the-art technologies in both computation and communication domains since both will jointly play an essential role in realizing the DT in smart industries. 
%The history of humanity keeps on changing with the arrival of innovative technologies, such as the arrival of electricity that completely reshaped agriculture, health, and numerous critical industries.
\begin{figure}
\centering
\includegraphics[width=0.95\linewidth]{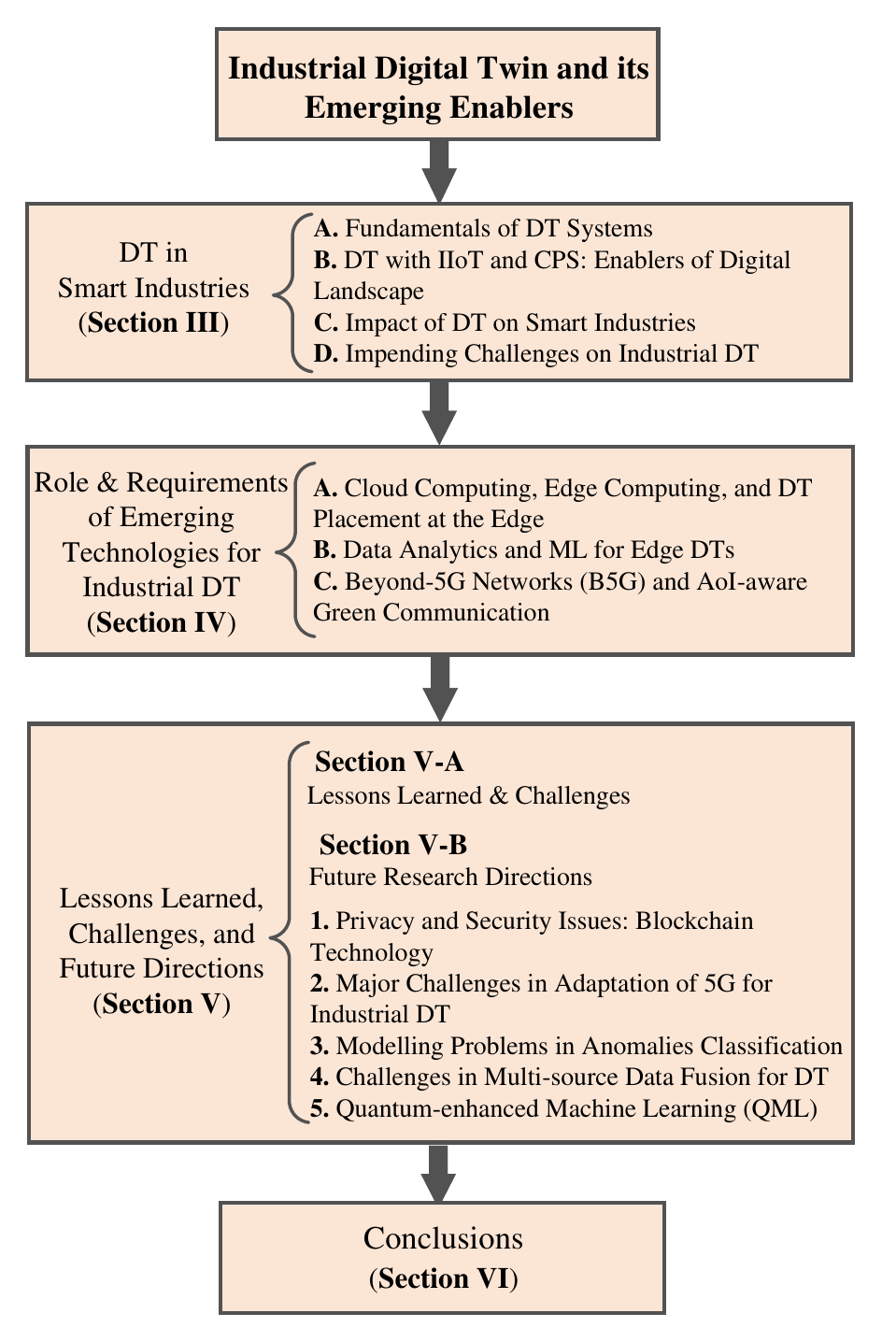}
	\caption{Structure and overview of our survey.}
	\label{fig:overview}
 	\vspace{-10pt}
\end{figure}
\subsection{Our Motivation and Contributions}
During the 2003-2011 time period, there was limited research on DT development due to the aforementioned reasons and technological constraints. Lesser publications are available on DT at the 2003-2011 timeline. However, on the other hand, other communication and computation enablers technologies, such as big data analytics, cloud computing, ML, and AI, continue seeing advances and exponential growth towards smart manufacturing. Moreover, the concept of DT was largely underestimated because of lag in the vision for DT significance, its adaptation, and long-term influence on real-time industrial applications. Nevertheless, this lag of vision changed when NASA in 2012 practically demonstrated the superiority of DT's adaptation in space flight shuttle program to solve the critical problem and devised a more specific definition. Since then, many DT applications in various fields have emerged, and the research academia has focused on it together with IIoT and CPS (as evident from Fig.~\ref{fig:publications}) due to many technological advancements in communication, sensing, and computation technologies. Keeping in view the current research trend and research gaps in DT adaptation, it can be argued that future research on DT and its practical deployment in the smart factories will experience exponential growth in the next 2-6 years. 

DT has already been adopted by various smart industries, complementing the vision of Industry~4.0. However, as the industry ecosystem's digital landscape embraces emerging technologies and tools, which include, but are not limited to, cloud and edge computing, ML and AI, age of information (AoI), and beyond-5G (B5G) network services, it brings up some critical questions. Especially, what is the role of various emerging technologies in enhancing futuristic smart industries' performance, and how these emerging technologies will reshape DT's usage in smart industries? Similarly, what are the vital requirements of different use cases that have to be fulfilled by DT in conjunction with these emerging technologies for realizing the Industry 4.0 vision? 

To the best of our knowledge, there is no prior work on DT for smart factories keeping in view the role and requirements of emerging technologies at various layers of the factory communication stack. 
Our key contributions in this review paper can be summarized as follows:
\begin{itemize}%[leftmargin=*]
\item We review the recent research on the use of DT in smart industries, elaborate upon functional aspects of DT, and highlight its appeal for smart industries. Moreover, we provided the taxonomy for DT usage in various industrial applications and identified the impending challenges in terms of communication and computation requirements for industrial DT.  
\item We discuss the current state-of-the-art developments in emerging technologies, especially the role of edge-cloud computing, ML and data analytic, federated learning, B5G/6G networks, green communication, and AoI, and their implications and significance on the performance of DTs. 
\item We discuss the DT placement strategies at different industrial communication layers to address the identified critical requirements. For instance, migrating DT capabilities from the cloud to the edge layer can address security, computation, and stringent quality of service (QoS) targets of factory floor applications.
\item Finally, we summarized the lesson learned from our thorough review work and outlined the possible future research opportunities and challenges in emerging technologies to facilitate DT's adoption in industries.
\end{itemize}
The rest of the article is organized as follows. Section~\ref{sec:IDTsec} gives an overview of the DT for smart industries, followed by Section~\ref{sec:EmergingTechnology} that discusses the role and requirements of emerging DT-enablers and technologies while focusing on B5G network services, AoI, ML, and mobile edge computing. Section~\ref{finalsection} discusses the future opportunities and challenges, and Section~\ref{sec:conclusion} gives concluding remarks. The overall structure of the article is given in Fig.~\ref{fig:overview}.
%%%%%%%%%%%%%%%%%%%%%%%
\section{DT in Smart Industries}
\label{sec:IDTsec}
\begin{figure*}[!t]
\centering
\includegraphics[width=0.7\linewidth]{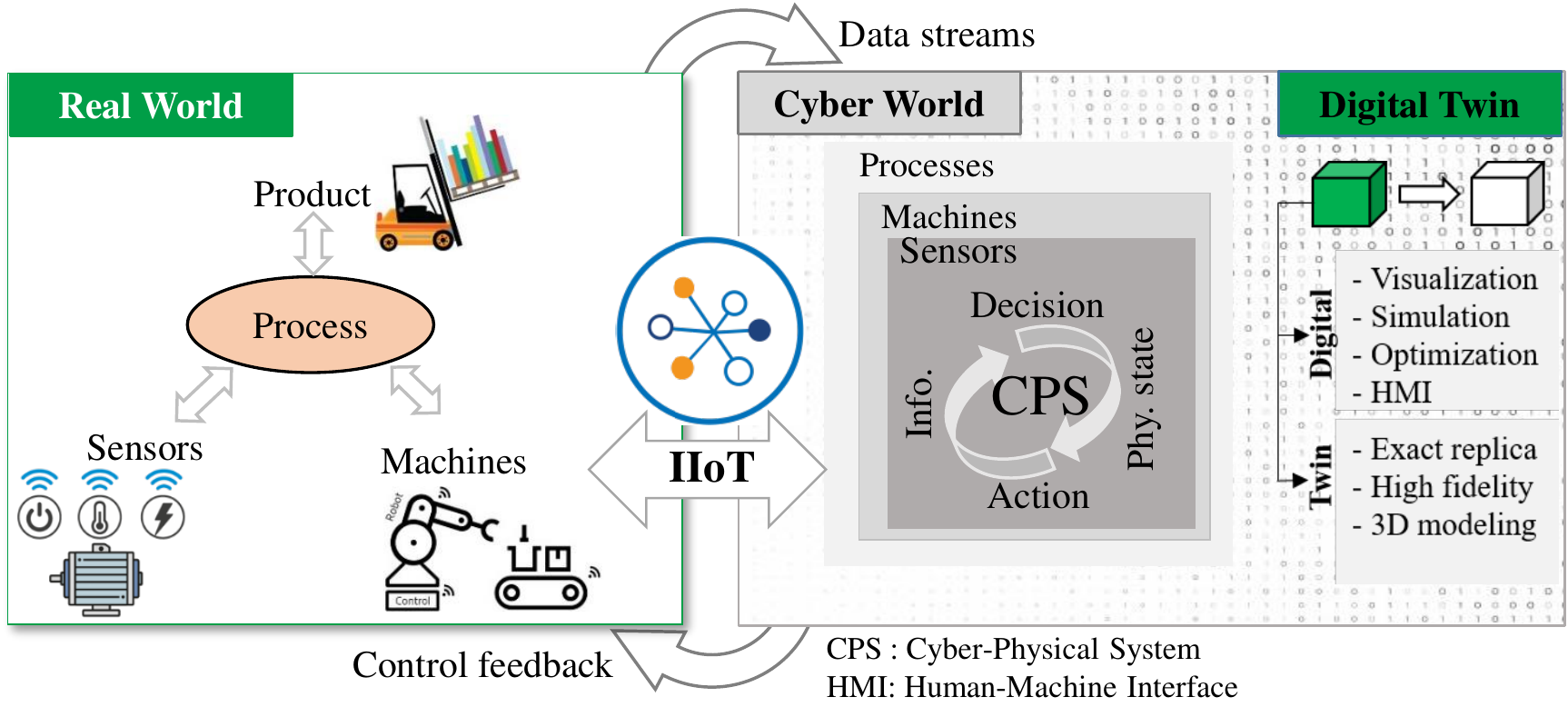}
	\caption{Mapping of real-world industrial objects and processes to the cyber-world based on IIoT and other data streams, with DT complementing as well extending the functional aspects of CPS.}
	\label{fig:DT_AM}
 	\vspace{-10pt}
\end{figure*}
This section provides an overview of DT fundamentals and the impactful role DT plays in tandem with IIoT and CPS inside the factory ecosystem to change the digital landscape. Furthermore, we classify the DT usage and its significance in numerous innovative industries and identify the critical challenges for the adoption of industrial DT.
\subsection{Fundamentals of DT Systems}
\label{sec:fundamentals}
A DT system of a smart industry forms a virtual image of physical objects in a factory environment, i.e., it depicts a \textit{living} digital simulation model of the physical counterparts in a factory, as shown in Fig.~\ref{fig:DT_AM}. A DT model is often confused with \textit{digital shadow} or \textit{digital model}; however, in the latter approaches, there are no automated exchange of control data between the image created in the virtual space and the physical objects to alter the industrial processes~\cite{kritzinger2018digital,ladj2021knowledge}. In contrast, the DT system of a single robotic machine or entire physical space of a factory continuously updates and evolves in real-time together with its physical counterpart to show the operating status, health conditions, and collaborating positions~\cite{madni2019leveraging,tao2017digital}.

To create a twin model of an object, integration of numerous communication technologies, cloud services, data analytics, and learning techniques is required~\cite{random3}. In this respect, the data sources for analytics and learning can be, for instance, individual sensors, similar machines in different systems, recorded data of faulty machines, and input of technical experts~\cite{random6,biesinger2019digital}. 
% such as ML and AI, cloud services, IIoT, spatial network graphs, and software data analytics~\cite{qi2018digital,CDT}. 
% The learning sources for DT can be from the individual sensors/machines data, the data from other similar machines in different systems, old recorded data from faulty machines, inputs from technical experts, and other fleets of machines in separate facilities~\cite{biesinger2019digital}. 
The inflow of information from all the sources significantly contributes to the development of agile and fast DT models, while the information is often stored in the cloud using dedicated network infrastructure.
%\vspace{-10pt}
% is stored in the cloud through a dedicated networking infrastructure and significantly contributes to the development of agile and fast digital models, i.e., DT of the smart factory.
\subsection{The Role of DT across Industries}
\label{sec:TwinIIoT}
The integration of IIoT and CPS with DT is critical in realizing intelligent factory machines since the high-value real-time data is generated throughout the working cycle of machines~\cite{DTandCPS}. Also, it enables machines to interact and evolve synchronously with other machines in cyberspace; thus allowing to assist and optimize various mission-critical applications in manufacturing and automation~\cite{seeboIoT2019}. 
The DT of intelligent machines recreates the factory ecosystem's physical space, enabling them to interact and evolve synchronously with other machines to assist and optimize various mission-critical applications, i.e., manufacturing and automation~\cite{seeboIoT2019}. 
% { \color{blue}
% Some of the blah blah blah (case study and taxonomy)
% \begin{itemize}
%     \item Manufacturing: \cite{manufact1}, \cite{manufact2}, \cite{manufact3}, \cite{manufact4}
%     \item Automobile: \cite{autom1} \cite{autom2} \cite{autom3} \cite{autom4}
%     \item Aerospace: \cite{aero1} \cite{aero3} \cite{aero3} \cite{aero4}
%     \item Windfarm: \cite{windf1}, \cite{windf2}, \cite{windf3}, \cite{windf4}
%     \item Healthcare: \cite{health1}, \cite{health2}, \cite{health3}, \cite{health4}
% \end{itemize}
% }
In Fig.~\ref{fig:DT_App_Enab}, we classified and referenced the latest case studies of DT usage reported in the literature for different innovative industries that come under the vision of Industry 4.0, i.e., manufacturing~\cite{manufact1,manufact2,manufact3,manufact4}, automobile~\cite{autom1,autom2,autom3,autom4}, aerospace~\cite{aero1,aero2,aero3,aero4}, windfarm~\cite{windf1,windf2,windf3,windf4}, and healthcare~\cite{health1,health2,health3,health4}. Moreover, Fig.~\ref{fig:DT_App_Enab} explicates the impact of valuable essential services provided by industrial DT in classified enabling application domain, and the subsequent subsection explores the vital impacts of industrial DT.
% Significant advantages of DT for smart industries are~\cite{munirathinam2020industry}:
%\label{sec:Digitaltwin}
\subsubsection{Data Visualization}
\label{sec:data}
%Visualization of mission-critical factory data enables the upright decision-making ability at factory management, albeit only human learning. 
% The complex process of machines in big manufacturing and automation industries is way advanced and sophisticated. It is not easy for technical management teams to make decisive actions on live-data presented in plain data-sheets and figures~\cite{tong2019real}. 
In industries, the manufacturing and automation processes are advanced and complex, thus making it nontrivial for technical and management teams to take decisive actions from the data in raw data-sheets and figures~\cite{munirathinam2020industry,tong2019real}.
% The use of DT plays its crucial part in bridging this gap of deep insight by integrating the visualization of the live data from machines in the virtual image/digital model of the whole factory. The redundant data from unwanted sources can be removed from the DT's visualized data to give a clear insight into the complex factory processes~\cite{zheng2019application}.
The DT bridges this gap by integrating the visualization of live data from machines in the virtual image or digital model.
% of the whole factory. 
Besides, any data redundancy can be removed from visualization to develop clear insight into complex factory processes~\cite{zheng2019application}.
Moreover, each deployed machine or robot's physical parameters, e.g., temperature, rusting, failure rate, and working conditions can be accessed. 
% , and the faulty machine can be easily identified. 
For example, a joint project by Altair, MX3D, and ABB showed a working DT model with visual settings for a 3D printed customized manufacturing robot~\cite{Altair2020}. The DT model of the robot and visual access to its time-series data has increased the robot's performance, which could be exploited to achieve higher precision and isochronous operation in smart factories.

\subsubsection{Collaboration at Management Levels}
\label{sec:collaboration}
% {\renewcommand{\arraystretch}{1.1}
% \begin{table}[t!]
% \centering
% 	\caption{Taxonomy of DT use cases for different industrial domain} 
% \begin{tabular}{|c|c|}
% \hline
%  \textbf{DT Application Domain} & \textbf{References} \\ \hline \hline
% \textbf{Manufacturing} & \cite{manufact1}, \cite{manufact2}, \cite{manufact3}, \cite{manufact4}
% \\ \hline    
% \textbf{Automobile} & \cite{autom1} \cite{autom2} \cite{autom3} \cite{autom4}
% \\ \hline    
% \textbf{Aerospace} & \cite{aero1} \cite{aero3} \cite{aero3} \cite{aero4} \\ \hline \textbf{Windfarm} & \cite{windf1}, \cite{windf2}, \cite{windf3}, \cite{windf4}
% \\ \hline 
% \textbf{Health} & \cite{health1}, \cite{health2}, \cite{health3}, \cite{health4}
%     \\ \hline
% \end{tabular}
% \label{taxonomytable}
% \vspace{-10pt}
% \end{table}
% }
Another crucial role of DT is to increase collaboration between the stakeholders, management authorities, expert teams, and the ground staff to monitor the facility output actively and weigh in if any input is required~\cite{lim2019state}. This collaboration provides the data scientists, field engineers, designers, and product managers, a deep insight into the complex processes of a manufacturing facility~\cite{macchi2018exploring}. Also, it gives a better comprehension of working knowledge, which helps design new prototype systems and test them quickly with increased efficiency. 
One example is ThyssenKrupp, a leading elevator manufacturer, which collaborated with Microsoft and Willow to built an intelligent cloud-enabled DT model for a 246-meter innovation test tower in Rottweil, Germany~\cite{Azure}. The collected data from hundreds of sensors, installed across the building, are integrated to create the building's digital replica in the cloud, giving a unique visual insight to perform asset and resource management in real-time. 
%It helps to understand the nature of building occupants, in terms of how they use the installed resources and what new resources should be added.
\subsection{Impact of DT on Smart Industries}
\label{sec:TwinImpact}
\vspace{-5pt}
%Impact of DT on Smart Industries: Manufacturing and Automation
%The worthiness of digital twin in the SI case can be judge by the fact that it adds significant value to the manufacturing and automation businesses throughout the entire lifecycle of operating machines in factory plants~\cite{lim2019state}. 
The impact circle of DT on the critical factories can be identified as,~\cite{gunasegaram2021towards,lu2020digital,tao2019make}:

\begin{enumerate}%[leftmargin=*]
\item \textbf{Product manufacturing and designing}: The availability of machines' DT enables accurate prediction of failure in the production process before affecting a plant's output targets. If system enhancement is desired, performance parameters can be adjusted and simulated in DT without imperiling the operation of the entire production.
%Then, the twin model's optimized approach is applied to real-time products in a manufacturing facility. 

\item \textbf{Field products}: 
It is more manageable to access and analyze the DTs for remote commissioning and diagnostics of deployed field products. It lowers service costs by remotely configuring faulty parts of a product, which can be ordered and replaced accordingly, for new customers.
% , thus increasing customer satisfaction. 
% Moreover, the DT's status can remotely identify the faulty parts of a product, which can be ordered and replaced accordingly.}
%In case of malfunction in the product, instead, the technician goes before checking the machine physically; the DT status at a particular time can identify the faulty part for a technician from a remote position, which can be ordered and replaced accordingly.

\item \textbf{Future products}: DT can predict machines' faulty behavior in complex systems, design newer and better systems from the learned history of machine operating conditions, and optimize a facility's efficiency and output.
%The pattern of customer usage with the performance of products over its lifecycle is recorded in the DT over time, giving better insight into the technical teams associated with the intricate process of new machine development and manufacturing.
\end{enumerate}
By catering to customer satisfaction and efficient working of smart factories, these DT-based end-services can undoubtedly increase the profit margin and market share of factory owners. For example, American electric power (AEP), which supplies electricity to more than 5 million customers, is developing a DT model of the US's most significant power transmission network with the specialized modeling and analysis software PSS{\textsuperscript \textregistered}ODMS from Siemens. It tightly integrates the electrical grid network with its virtual twin model~\cite{siemensgrid}. Otherwise, the grid network planning and provisioning of services to customers were becoming complicated with traditional (manual) methods of sharing the technical data among the various utility systems.
Similarly, ABB's state-of-the-art electromagnetic (EM) flow measurements products integrate DT technology to build up the predictive model of EM flow during production processes using multiphysics finite element analysis (FEA) techniques~\cite{ABB}. In particular, DT usage mimics the virtual EM flow process, giving visual insights to acquire performance complexities.
\begin{figure}[!t]
\centering
\includegraphics[width=1\linewidth]{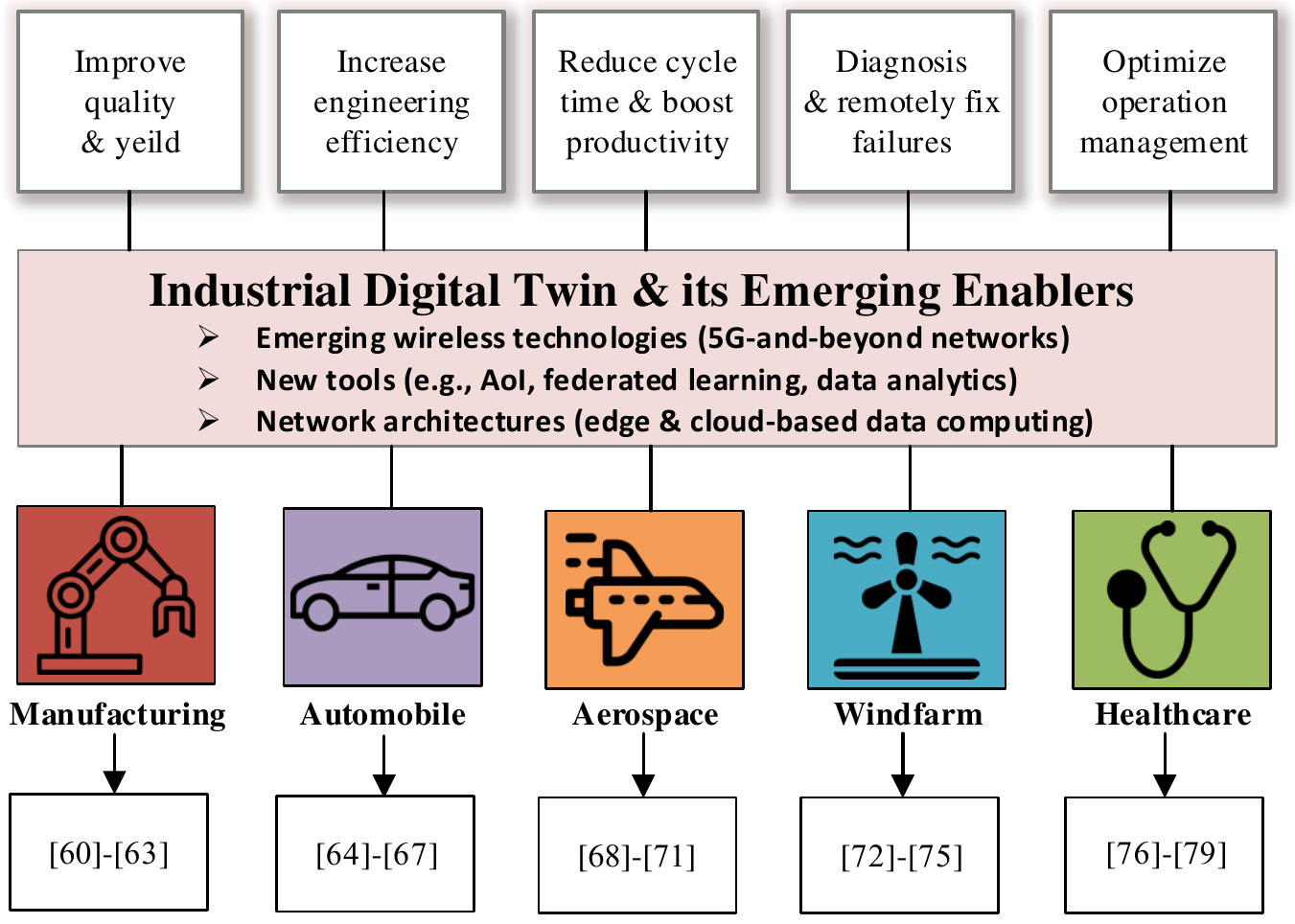}
	\caption{An illustrative block diagram depicting the significance and impact of an industrial digital twin along with its critical communication enablers on numerous classified innovative industries}
	\label{fig:DT_App_Enab}
  	\vspace{-10pt}
\end{figure}
%Integrating both DT and cloud capabilities with the installed IIoT at different smart grid nodes of the network in 11 states allows them to have a centralized data approach among all company utility domains on a single page. Thus, it empowers the company to have improved asset control with simplified planning and the capability of seeing into the future needs of its customers and satisfaction with current services.
\subsection{Impending Challenges in Industrial DT}
The initial coined idea of DT was in the context of increasing the product life cycle of an industrial machine and learning from the anomalies and malfunction over time, which tends to design it better.
% Initially, the coined idea of DT was from the context of increasing the product life cycle of machine/device in various applications and learning from the product anomalies and malfunction, which trends to design it better. 
However, the simultaneous interplay of smart industry twin's with all emerging communication and computation technologies in large-scale factory scenarios inherits significant challenges and hurdles (c.f.~Fig.~\ref{fig:DT_App_Enab}).For example, 
\begin{itemize}%[leftmargin=*]

\item A large amount of data from numerous factory floors needs to be transmitted for mapping a large number of industrial devices with their virtual counterpart in the cloud or possibly at the edge, while the communication resources are limited.
\item The communication burden caused by this frequent real-time interaction of factory floor machines with the DT residing in the cloud may lead to intolerable delays for time-critical applications.
\item The addition of edge architecture with the cloud brings new roles and adjustments to a digital twin's deployment strategies to address the requirements on performance metrics, such as big data management, communication latency, reliability, packet loss ratio (PLR), data update, data size, security, and privacy. 
\item The massive inflow of incurred machine data from the factory manufacturing floor using communication infrastructure requires enhanced raw data preprocessing and the latest computation-efficient data analytics and learning techniques to build up the industrial DT.
\item The energy constraints and AoI requirements set by the applications' requests limit the collected data update rates in meeting the goal of energy-efficient green communication for industrial devices while satisfying the information freshness at the industrial DT.
\end{itemize}

% \begin{comment}
% --R3 (4) CDT and EDT play different roles in IIoT. The authors should elaborate on the differences between CDT and EDT
% %%%%%%%%%%%%%%%%%%%%%
% \end{comment}
\section{Role and Requirements of Emerging Technologies for Industrial DT}
\label{sec:EmergingTechnology}
\begin{figure*} 
\begin{center}
  \includegraphics[width=0.75\linewidth]{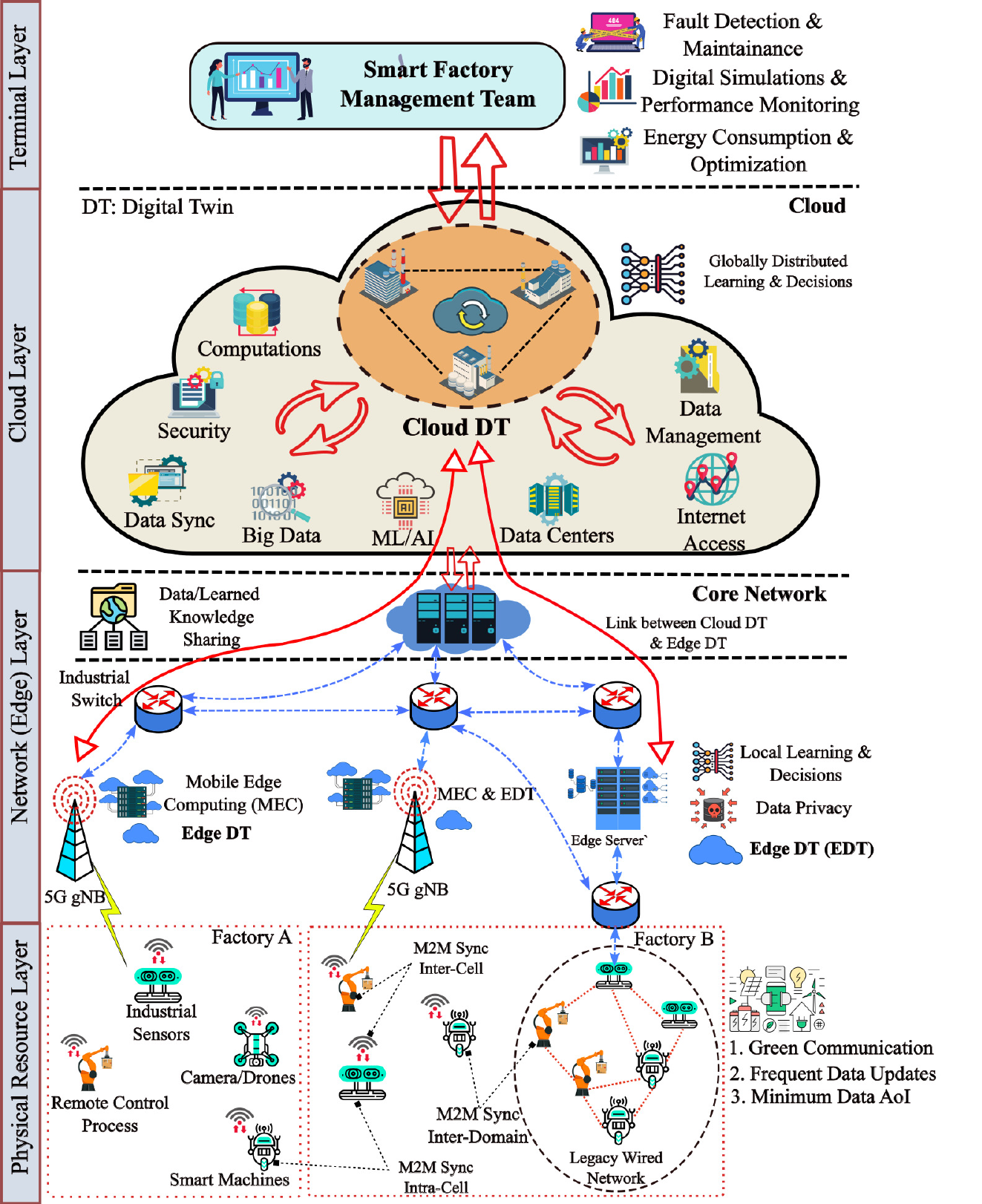}
  \caption{Illustration of B5G and cloud/edge-based DT layered architecture for smart industries. Main ideas: a) a part of CDT is shifted to the edge layer to make local learning and make decisions quickly (federated learning), b) EDTs are developed at the edge layer, i.e., at 5G gNodeB (gNB) or edge server, which takes the inflow of data from numerous sources, computes and locally learn, and c) 5G gNB provides the computation-enabled NextG network services to provide efficient and reliable wireless connectivity for the factory devices.}
  \label{maindiagram}    
  \vspace{-15pt}
\end{center}
\end{figure*} 
%In addition to this, a new service class of advanced, flexible, and efficient DT has emerged that complements the vision of adequate smart services to the consumer market especially in smart manufacturing use cases~\cite{tao2017digital}. 
The integration of DT with the emerging technologies, i.e., edge layer architecture, B5G network services, state of the art ML and AI frameworks, can open up many new potential use cases of DT and accelerate the digital transformation of smart industries. Table~\ref{table:I} summarizes the various critical requirements of industrial use cases, and it is required to maintain these demands by the DT of a smart factory. Note that each case's generated data group has a class of \emph{data and big data} category, which is not mentioned in the table. The \emph{data update time (msec)} applies to the periodic updates of event-based or sporadic data traffic generated.
%to utilize the enabling technologies efficiently. 
These emerging technologies are explored in subsequent subsections for their adoption in DTs, and their roles and needs are also discussed at each layer of the communication stack, as identified in the smart factory scenario in Fig.~\ref{maindiagram}. 
%Initially, the coined idea of DT was from the context of increasing the product life cycle of machines/devices in various applications and learning from the product anomalies and malfunction trends to design it better. However, DT's integration for smart industries in the new space of different emerging technologies has given rise to a new service class of advanced, flexible, and efficient DT. Table.~\ref{table:I} summarizes the various critical requirements of smart industries use cases, and these requirements must be fulfilled by these new classes of DT to utilize the enabling technologies efficiently. These emerging technologies are explored in the subsequent subsections from the DT usage perspective and discuss their roles and needs at each layer of the communication paradigm, as identified in the smart factory scenario of Fig.~\ref{maindiagram}.
%*********************************
\subsection{Cloud-/Edge-Computing and Industrial DT Deployment}
% %\label{Edgecomputing}
% \subsection{Cloud Computing and Cloud-based Digital Twins (CDTs)}
% \label{cloudcomputing}
Cloud-/Edge-Computing (cloud-edge computing) is a critical component of Industry 4.0 to ensure on-demand availability of high computing resources, e.g., as shown for aerospace manufacturing industry in~\cite{8850990}, vehicular intelligence towards connected smart vehicles~\cite{zhang2021adaptive}.
%, such as authors in \cite{8850990} uses combination of ML and cloud computing technologies in aerospace manufacturing industry to enable construct of DT. 
The vital strengths of high computing power, massive data storage capacity, data analytics, service-oriented architecture with a sizeable autonomous structure have led to a massive adoption of cloud computing in today's smart industries~\cite{coelho2021automatic}. Numerous CPS- and IIoT-based machines generate a large amount of data during the intricate manufacturing process, which has to be transferred and stored in the cloud~\cite{Cloud1}. By this, the industries can reduce the cost of dedicated data centers, which also brings global access and management to factories~\cite{cloud2}.

\subsubsection{Cloud-based Digital Twins}
The creation of a virtual digital image of a factory from the inflow of data from heterogeneous sources in the cloud leads to a significant class of twins, termed as cloud-based digital twins  (CDTs)~\cite{cloud4}.
Fig.~\ref{maindiagram} shows the cloud-native CDT service closely integrated with the upper factory management layers. In the cloud, necessary operations, e.g., pre-processing of machine data and big data analytics, are applied for efficient data management and utilization~\cite{cloud3}. By using that, CDT brings more possibilities; it enhances the collaboration and visualization for intelligent decision making, in addition to the advantages discussed in Section~\ref{sec:TwinIIoT}. 
%Moreover, the complex network of machines and smart devices in factories are relieved of dedicated infrastructure for data storage and computation systems. 
Moreover, CDT allows the training of a complex network of all industrial assets
%machines, smart devices, and control systems 
with high power computing (HPC), deep learning (DL) and AI. 
\definecolor{Gray}{gray}{0.9}
\definecolor{Gray1}{gray}{0.7}
\bgroup
{\renewcommand{\arraystretch}{1.3}
\begin{table*}[ht]
\centering
	\caption{A Summary of smart industries requirements in Industry 4.0 (based on~ \cite{shahzad2020internet,ho2019next,schulz2017latency,yan2017industrial,5GAcia}). 
% 	Moreover, the \emph{data update time (msec)} applies to the periodic updates of event-based or sporadic data traffic generated. 
}
\scalebox{0.93}{
\begin{tabular}{|c|c|c|c|c|c|c|c|c|}
\hline
\begin{tabular}[c]{@{}c@{}}\textbf{Smart Industries}\\ \textbf{(Use-cases)}\end{tabular} & \textbf{Security} & \begin{tabular}[c]{@{}c@{}}\textbf{Data Size}\\ \textbf{(bytes)}\end{tabular} & \begin{tabular}[c]{@{}c@{}}\textbf{Device}\\ \textbf{Density}\end{tabular} & \begin{tabular}[c]{@{}c@{}}\textbf{Latency}\\ \textbf{(msec)}\end{tabular} & \begin{tabular}[c]{@{}c@{}}\textbf{Availability}\\ \textbf{(\%)}\end{tabular}  &
\begin{tabular}[c]{@{}c@{}}\textbf{Reliability}\\ \textbf{(PLR)}\end{tabular} &
\begin{tabular}[c]{@{}c@{}}\textbf{Data Update}\\ \textbf{Time (msec)}\end{tabular} & \begin{tabular}[c]{@{}c@{}}\textbf{Communication}\\ \textbf{Range}\end{tabular} \\ \hline\hline
\rowcolor{Gray}
\begin{tabular}[c]{@{}c@{}} Factory Manufacturing\\ Cells\end{tabular}
   & Yes & $<$20 & \begin{tabular}[c]{@{}c@{}} 0.33-3\\devices/$\textrm{m}^2$ \end{tabular} & 4 & $>$ 99.9999 & $10^{-9}$ & 40-50 & 60-120 m  \\ \hline

\begin{tabular}[c]{@{}c@{}} Robots in Assembly\\ Process\end{tabular}   
   & Yes & 40-240 & \begin{tabular}[c]{@{}c@{}} 0.33-3\\devices/$\textrm{m}^2$ \end{tabular} & 3-9 & $>$ 99.9999 & $10^{-9}$ & 2-10 & 60-120 m \\ \hline
\rowcolor{Gray}
\begin{tabular}[c]{@{}c@{}} Camera-controlled \\Remote Operation \end{tabular}     
   & Yes & $<$3K  & \begin{tabular}[c]{@{}c@{}} 0.33-3\\devices/$\textrm{m}^2$ \end{tabular} & 8-95 & $>$ 99.9999 & $10^{-9}$ & 25-40 & 60-120 m \\ \hline
 
\begin{tabular}[c]{@{}c@{}} Factory Machines in \\ Printing \end{tabular}      
  & Yes & 25-35 & \begin{tabular}[c]{@{}c@{}} 0.33-3\\devices/$\textrm{m}^2$ \end{tabular} & 2 & $>$ 99.9999 & $10^{-9}$ & 1-2.5 & 60-120 m \\ \hline
\rowcolor{Gray}
\begin{tabular}[c]{@{}c@{}} Factory Machines in \\Packaging  \end{tabular}     
 & Yes & 30-50 & \begin{tabular}[c]{@{}c@{}} 0.33-3\\devices/$\textrm{m}^2$ \end{tabular} & 1 & $>$ 99.9999 & $10^{-9}$ & 4-6 & 60-120 m \\ \hline
 
 \begin{tabular}[c]{@{}c@{}} Motion Control in \\Isochronous Robots \end{tabular}   
  & Yes & 50-260 & \begin{tabular}[c]{@{}c@{}} 0.33-3\\devices/$\textrm{m}^2$ \end{tabular} & 1 & $>$ 99.9999 & $10^{-9}$ & 0.5-2 & 60-120 m \\ \hline
 \rowcolor{Gray}
 \begin{tabular}[c]{@{}c@{}} Machine Tools at \\ Factory \end{tabular}   
  & Yes & 40-60 & \begin{tabular}[c]{@{}c@{}} 0.33-3\\devices/$\textrm{m}^2$ \end{tabular} & 0.5 & $>$ 99.9999 & $10^{-9}$ & 0.5-1 & 60-120 m \\ \hline
 
 \begin{tabular}[c]{@{}c@{}} Monitoring Process \\(Factory Automation) \end{tabular}  
  & Yes & Varies & \begin{tabular}[c]{@{}c@{}} 10000\\devices/plant \end{tabular} & 45 & 99.9 & $10^{-3}$ & 80-4500 & 150-600 m \\ \hline
  \rowcolor{Gray}
  \begin{tabular}[c]{@{}c@{}} Remote Control Process \\(Factory Automation) \end{tabular} 
 & Yes & Varies & \begin{tabular}[c]{@{}c@{}} 10000\\devices/plant \end{tabular} & 45 & 99.99 & $10^{-5}$ & 80-4500 & 150-600 m \\ \hline
 
 \begin{tabular}[c]{@{}c@{}} Grid Stations High Voltage \\(Smart Grid) \end{tabular} 
  & Yes & 100-1100 & \begin{tabular}[c]{@{}c@{}} 1000\\devices/$\textrm{km}^2$ \end{tabular} & 6 & 99.999 & $10^{-6}$ &5-100  &\begin{tabular}[c]{@{}c@{}} Few meters to\\kilo-meters \end{tabular} \\ \hline
  \rowcolor{Gray}
 \begin{tabular}[c]{@{}c@{}} Medium-Low Volatge at \\Transmission Lines \\(Smart Grid) \end{tabular}  
  & Yes & 100-1100 & \begin{tabular}[c]{@{}c@{}} 1400\\devices/$\textrm{km}^2$ \end{tabular} & 20 & 99.9 & $10^{-3}$ & 5-100 & \begin{tabular}[c]{@{}c@{}} Few meters to\\kilo-meters \end{tabular} \\ \hline
 
\end{tabular}}

\label{table:I}
% \vspace{-5pt}
\end{table*}
}
\egroup
% \subsection{Edge Computing and DT Placement Strategies}
% \label{Edgecomputing}
\subsubsection{Emergence of Edge-based Digital Twins}
\label{sec:E-DT}
CDTs have certain inherent limitations of cloud architecture for stringent time-critical industrial communications, e.g., high round-trip time (RTT) with regular periodic data updates and end-to-end (E2E) latencies to the cloud~\cite{edge1}. 
Similarly, factory machines' reliability factor can drastically reduce with the outdated decisions for the critical sporadic events happening at the factory floor~\cite{edge2}.
%CDTs have certain inherent limitations in adapting emerging technologies for stringent time-critical applications, e.g., high round-trip time (RTT) with regular periodic data traffic updates and end-to-end (E2E) latencies to the cloud~\cite{edge1}. Similarly, factory machines' reliability factor can be drastically reduced due to the wrong decisions for sporadic events happening at the factory physical layer~\cite{edge2}. CDT needs a continuous stream of data from all periodic and sporadic events to learn and evolve with time. 
What if the DT in the cloud is deployed or shifted towards the factory network's edge layer, i.e., at the factory gateways, industrial controllers, cluster of machines, 5G gNB.
This emerging new cloud computing architecture named "edge computing" can address these drawbacks and brings new novel analytics and control strategies at the network edge. 

\subsubsection{DT Deployment at the Edge for Critical Communications}
\label{sec:Requirements}

The edge servers at the factory network can take data readings from physical entities locally, store and pre-process it, make advanced computations, and have cloud-assisted analytics and real-time control~\cite{edge3}. Moreover, the edge network's end nodes (i.e., IIoT and CPS-based machines) have developed small-scale computation power over time~\cite{edge4}. These computing resources at the underlaying edge architecture can bridge the gaps for a new class of smart vertical industries in tandem with cloud computing. The edge twins can be independently created locally from the heterogeneous streams of incoming data, or a copy of the CDT model can be provided at the network edge. The CDT continuously gets updates from the local edge-based digital twins (EDT) that is running close to the factory physical layer, as shown in Fig.~\ref{maindiagram}. In either case, EDT brings flexibility and agility to the decision-making process for critical events, i.e., insight for performance optimization in machine processes, abrupt anomalies, and disaster situations.

Table~\ref{table:I} shows that security, latency, and reliability are critical requirements for smart manufacturing, smart grids, and intelligent vehicular domains. Bringing DT capabilities from the cloud layer to the edge devices or servers indeed reduces the impact of latency and decision reliability as it lessens the cloud dependency by making the critical decisions locally at the EDTs.
%Moreover, the continuous transmission of big data of the factory to the cloud for storage and analytics services can be costly. The data pre-processing at the edge devices reduce the large volume of data transmission to the cloud. There is a growing vulnerability in cloud-based applications' security due to the increase in data breaches in cloud services. 
Moreover, while continuous transmission of big data from factory to cloud can be costly and vulnerable to data breaches, edge-based pre-processing can reduce such concerns. In \cite{edge5}, the authors addressed the security requirements of users' data in edge computing by integrating the DT and blockchain technology at the edge layer, which increases the robustness of the IoT networks.
The computation of sensitive factory data can be performed at the edge layer to facilitate EDT. In the event of disconnection from the cloud, analytics can still run at the edge device, keeping the real-time continuous self-learning and evolving EDT with time. EDT can update the CDT once the connection restores, thus increasing the resilience of the smart industry network.
%*********************************

\subsection{Data Analytics and ML for EDTs}
\label{ML}
ML- and DL- algorithms learn and approximate a mathematical model based on the set of provided sample data, and can help in predicting disaster events and anomalies while bringing intelligence to various DT-based applications~\cite{7548905,yang2021application}. 
%At the core, numerous data analytics techniques in the cloud or edge that aids the digital twins of factories to run on these ML algorithms that gather and pre-process the incoming raw data from various data sources~\cite{8863728}. 
At the cloud/edge, numerous data analytics techniques, with ML algorithms' help, pre-process the incoming multi-heterogeneous raw data in the cloud or edge that aids the Industrial DTs~\cite{8863728}.
Modern trends in the field of ML can enhance the usage of CDTs and EDTs across multiple industries. 
\subsubsection{Data Sources, Pre-processing, and Data Fusion for Analytics}
\label{sec:DataSources}
Application of data analytics framework on a continuous stream of incoming time-series factory data plays an essential role in the perpetual update of the DT at both cloud and edge~\cite{Data1}. In smart industries, generated data can be classified into two categories based on the source of their origination at the physical layer, i.e., \textit{factory field data} and \textit{factory management data}~\cite{Data2,qi2018digital}. 
\begin{enumerate}%[leftmargin=*]
\item \textbf{Factory field data} is composed of multiple data inflows from the physical layer of an operating factory. For example, environmental data related to air quality, temperature, humidity, and other essential data linked with machine performance is collected from IIoT and CPS. 
\item \textbf{Factory management data}, carrying information on product planning, design schematics, service management, and finance, originate from the numerous information and computer-aided systems, such as manufacturing execution system (MES), enterprise resource planning (ERP), computer-aided design (CAD), and computer-aided engineering (CAE).
\end{enumerate}
This inflow of data at both the edge and cloud layers forms the building block for realizing and updating CDTs and EDTs. However, the underlying physical layer's raw data is barely useful because of the multi-source and multi-scale, heterogeneous, and highly noisy data nature~\cite{Data3}. Hence, pre-processing of the data is needed before any ML-based analytics operation is applied to extract the valuable information for efficient simulation of DT at the edge and cloud layer. Moreover, data fusion techniques can be applied during the pre-processing step, where data from multiple data sources are fused for constructing accurate and reliable insights~\cite{Data4}. 
\subsubsection{Streaming Analytics for EDTs}
\label{sec:Analytics}
Traditional data analytics store the data first and then analyze it to extract insightful data patterns. 
%In the new streaming analytics model, incoming data is continuously analyzed simultaneously as the factory events, generating the time-series data, is still ongoing at the physical layer~\cite{SA1}. 
In the new streaming analytics model, incoming time-series data are continuously analyzed while the machine processes are still in progress at the factory floor~\cite{SA1}.
Afterward, the processed data is stored for batch analysis. 
Moreover, traditional analytics at cloud and edge needs to store data first before any further analysis. However, as discussed in Section~\ref{sec:E-DT}, CDTs can induce large RTT latencies. The synergy of both streaming data analytics and edge architecture increases the agility in EDTs to address the stringent low-latency and mission-critical events~\cite{SA2}. Moreover, this synergic mode leads to better and faster insights at the EDTs to act locally on critical events and make the all-important decisions. 
\subsubsection{Emerging Trends in Machine Learning}
\label{sec:MLTrends}
 %ML and the newest deep learning(DL)-based frameworks are built in the cloud and edge to digitally model and classify the performance parameters, based upon which twins of the underlying physical objects or entire facility are derived. 
 ML- and DL-based frameworks are typically built in the cloud and edge layers to model and classify the performance parameters from industrial data, which are used to update the DT.
 However, the nature of time-series industrial data, originating from various machine processes, is different; it has large volume and dimensionality, and varying degrees of correlation and sensitivity depending on the time cycle~\cite{ML1}.
%  However, the nature of time-series industrial data originated from various machine processes is different as the data have large volumes, larger dimensions, collected at various time cycles with high sensitive value, and the degree of correlation between time series data varies~\cite{ML1}. 
 Hence, conventional ML techniques, such as regression and classifying techniques, cannot be applied. 
 %New emerging approaches in the ML area need to be explored for addressing the EDT requirements of low computation and better control on insights, and ensure sensitive industrial data security and critical communication needs of Table.~\ref{table:I}. 
 Therefore, new emerging ML approaches need to be explored for meeting the EDT requirements of low computation and better control on insights, while ensuring security and communication needs (see Table~\ref{table:I}).
For example, for anomalies detection in collaborative machines and safety precautions in factories, visual perception sensors like industry-grade video cameras are installed, which continually produce a time-series visual data~\cite{ho2019next}.
A separate powerful class of artificial neural networks (ANN), called convolution neural networks (CNN), has extensive usage in computer vision (CV) and perform better, especially on a cameras-originated perceptual data class that has an inherent property of local relationships among spatial dimensions inside images~\cite{ML2,coelho2021automatic}. 
Generally, CNNs had two parts: 1) feature extractors that learn features from raw data, and 2) trainable multilayer perceptron (MLP), which performs classifications based on input from learned features. However, it is noted that traditional ANN lags the support for performing spatio-temporal analysis on time-series as it does not use the past historical observations and information acquired in the previous steps of the learning/training process. For this purpose, various causal convolutional filters of CNN units are designed and utilized to use past information for learning long-term correlation in time-series data for accurate prediction~\cite[Chap.~3]{brownlee2018deep}. CNN applied on the perceptual data at the edge has the potential of continuously updating the EDT in real-time to detect and respond to anomalies appropriately. Similarly, recurrent neural networks (RNNs) and their extensions, i.e., gated recurrent unit (GRU) and long short-term memory (LSTM)-based neural networks, has also an inherent property of modeling past historical observations and spatio-temporal analysis on incurred time-series machine data for prognosis and forecasting applications~\cite{ma2020deep}.
\begin{figure*} 
\begin{center}
  \includegraphics[width=0.8\linewidth]{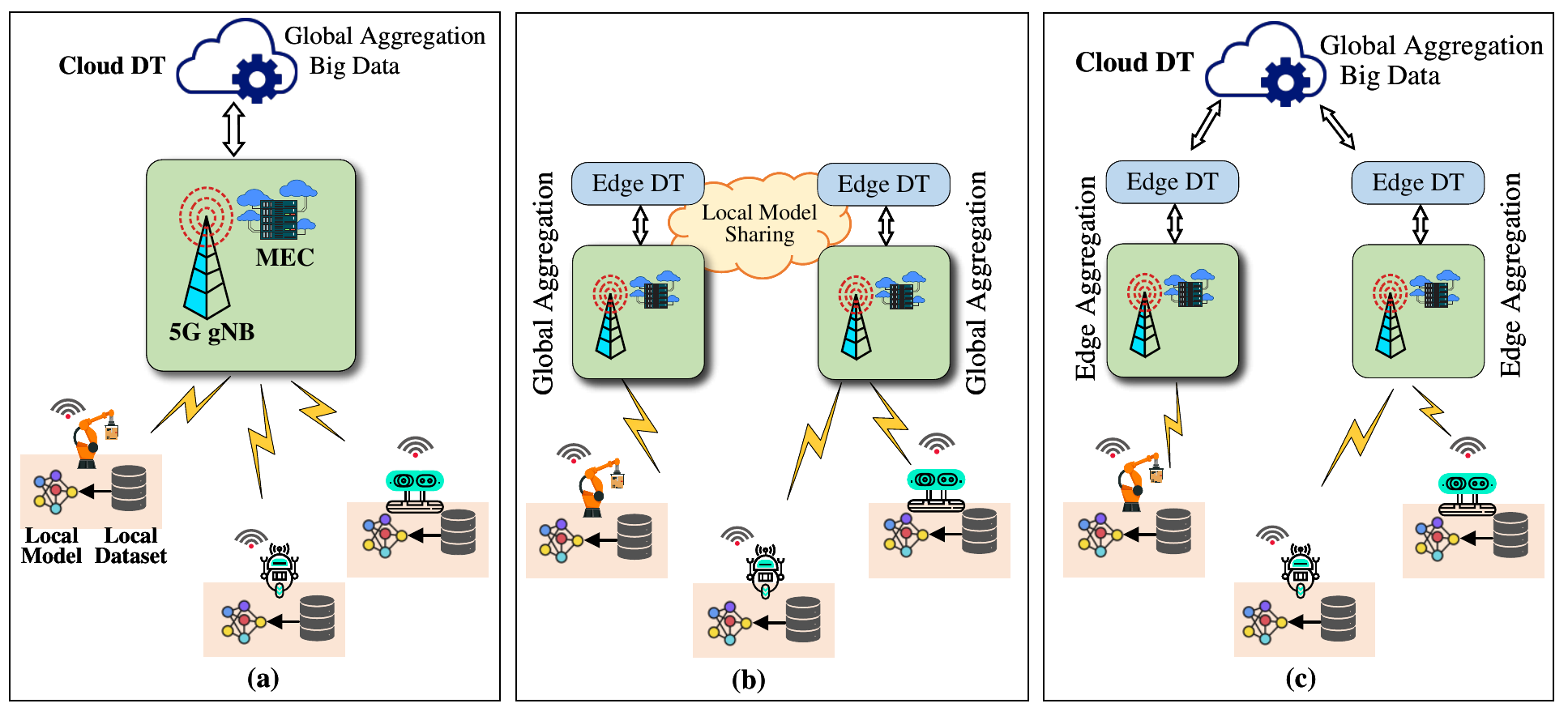}
  \caption{Federated learning methods based on: (a) centralized aggregating and computing approach, (b) decentralized aggregating and computing approach, and (c) fully-distributed aggregating and computing approach.}
  \label{federatedlearningplus5G}    
  \vspace{-15pt}
\end{center}
\end{figure*} 

However, the ML algorithm running on a single computing node with a centralized infrastructure will be insufficient for the multi-heterogeneous and enormous volume of generated data in factories~\cite{ML3}. Increasing resources on a single computation machine or complexity of DL frameworks by adding more fully connected hidden layers of neurons to learn all the performance parameters is not a go-to option for industrial big data. A different hierarchical approach of computing-based ML ecosystem can be adopted, which varies on the type and degree of hierarchical distribution. These include:
\begin{itemize}
    \item \textbf{Decentralized ML Computing} approach, in which various computing edge servers share the data set with its connected edge servers (peers) for computations, and no single master node (i.e., cloud) has centralized control over ML computations.  
    \item \textbf{Fully Distributed ML Computing} approach, where the master node makes the big data-based ML computational task by sharing the data set among the connected peers with having a single control over them~\cite{ML4}.
\end{itemize}
The discussion in Section~\ref{sec:E-DT} leads to a vital lesson that moving DT from the cloud to edge can bring many novel applications in smart industries. However, the lack of an efficient ML framework and a large volume of generated data inflow incurs high computations requirements for which distributed ML approaches serve the purpose. The core of this approach employs a parallelization technique, \textit{data parallelism}, which is applied to the partitioned machine data~\cite{ML5}. In data parallelism, the industrial data is initially split into multiple data cells at a cloud that shares one cell to the connected computational nodes through networked communication. Then, each node performs training, learning the optimized parameters, and transfers them among the nodes to update their learned parameters until they reach consensus on the learned parameters and submit it back to the cloud.
\subsubsection{Federated Learning Approach}
\label{sec:Federated Learning}
Data parallelism-based distributed ML approach addresses the efficient management of large ML computations at the cloud and edge. However, the potential risk of data breaches and inducing large latencies remains large as the sensitive data is shared by the master node (cloud) to computational devices on edge servers and slow updates to the DT at the factory layers~\cite{chen2021digital,Fed1}. Another ML approach, called "\emph{federated learning}" can be used, which utilizes a \textit{model parallelism} technique instead of \textit{data parallelism}. In model parallelism, the learned ML model or framework is shared with the computational edge nodes without exchanging local data by the master node~\cite{Fed2}. The master node chooses the ML framework for training and transmits it to the edge devices for training separately on the locally generated data of field devices without exchanging any local data. All edge devices share the optimized trained ML frameworks with the cloud, which pools the received models result and selects the best global model for further usage. This approach can address the data-security related issues and provide the real-time continuous learning evolution at the twins at both layers of cloud and edge. 
Fig.~\ref{federatedlearningplus5G} shows the various federated learning methods for CDT and EDT implementation at the factory floor based on the hybrid approach of model aggregation and computing techniques. A DT-based edge architecture is proposed and analyzed for IoT network in~\cite{Fed3}, which develops and trains the twin model on the devices' data using federated learning.
%\cite{Fed3} proposed and analyzed a DT based edge architecture for IoT network that developed and trained the twin model on the devices' data over time using a federated learning approach. 
Results from \cite{Fed3} show that real-time optimization of resource allocation to network is achievable using federated learning, even without uploading data to the cloud.
%Results from \cite{Fed3} showed real-time optimization of resource allocation to network is achievable, and the inclusion of federated learning leads to avoidance of submitting data to a centralized cloud node.
%%%%%%%%%%%%%%%%%%%%%%%%%%%%%%%
\subsection{5G-and-Beyond/6G Networks and AoI-aware Green Communication}
\label{5GRole}
Future digital industries need a service-based B5G/6G wireless network capable of: a) satisfy the stringent mission-critical communication requirements of Table~\ref{table:I}, and b) optimizing the radio and core network resource allocations for the diverse factory floor services.~\cite{B5G1,saad2019vision}. Integrating DT with 5G networks leads to a simulated end-to-end software replica of the underlying industrial network~\cite{B5G2,commMag1}. It can ensure the critical communication requirements for factory floor of Fig.~\ref{maindiagram} (\textit{Physical Resource Layer}) by having continuous analysis, predictions, and recommendations to provide hybrid 5G network services.
\subsubsection{Industrial B5G/6G Wireless Connectivity and Services}
\label{sec:5GNWservices}
\begin{figure*} 
\begin{center}
  \includegraphics[width=0.8\linewidth]{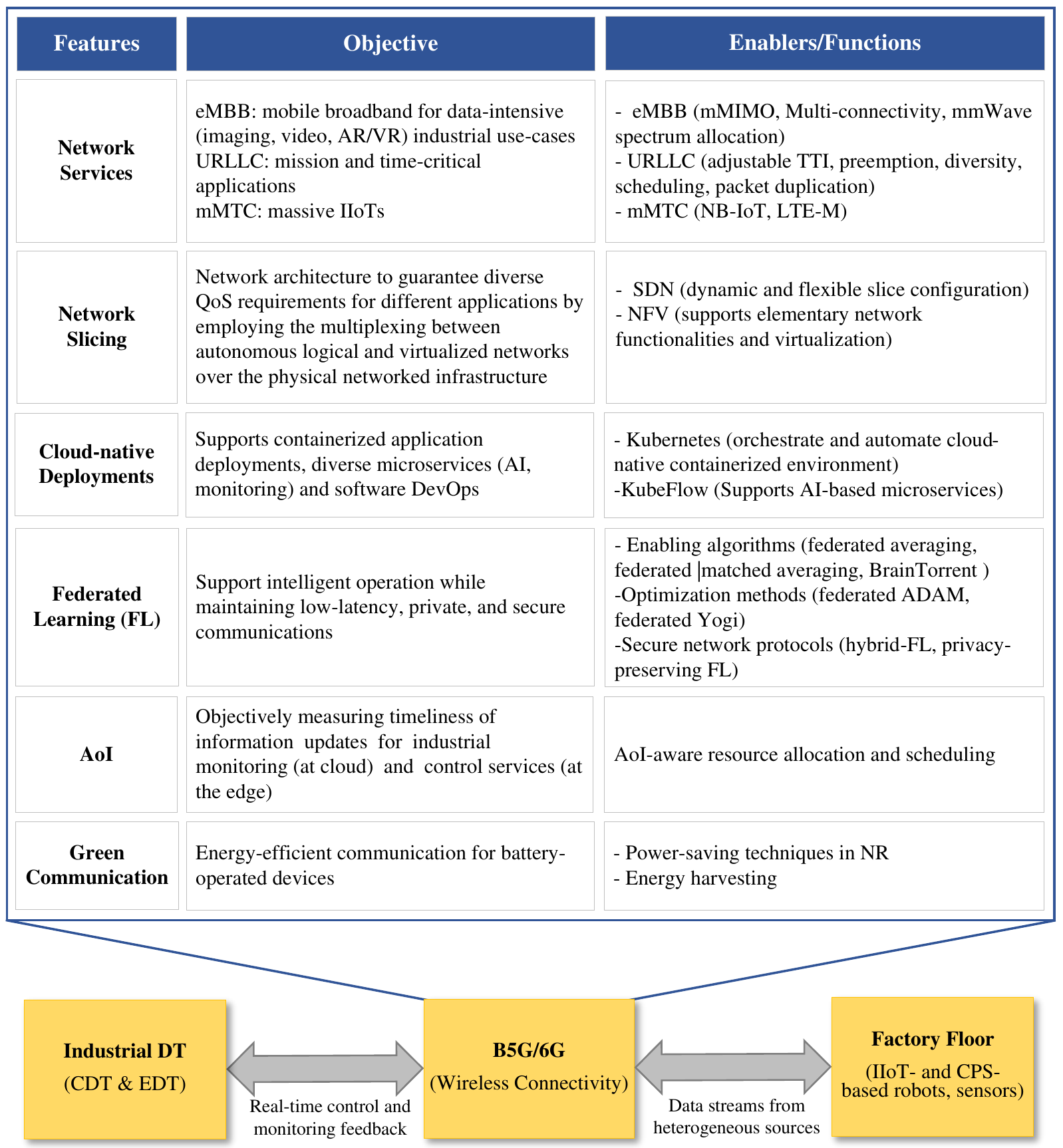}
  \caption{Features expected from B5G/6G in terms of industrial DT usage, objective and what technology enables these features.}
  \label{5G}    
  \vspace{-15pt}
\end{center}
\end{figure*} 
5G-and-beyond wireless networks are constantly evolving to provide service-based wireless access
to the futuristic industries with services as ultra-reliable and low latency communication (URLLC), enhanced mobile broadband (eMBB), and massive machine-type communication (mMTC)~\cite{cheng2018industrial}.
Similarly, the futuristic vision of 6G network design has been stepping forth based on the amalgamation of trending ICT and data technology in tandem with the cloud-native computing model to the 5G core network, increasing the prospects of new services.~\cite{zeb2021edge,AIref}. New features and enablers expected from B5G/6G networks for industrial DT are identified and summarized in Fig.~\ref{5G}.

Unique to 5G-and-beyond networks is the all-out effort from telecom standardization bodies (e.g., ETSI, 3GPP), regulators, service providers, operational technology (OT) companies, and manufacturers to transfer the technological advancements to industrial domain. In particular, key industry bodies from the manufacturing sector, which are the market representative to 3GPP, like 5G automotive association (5GAA), 5G alliance for connected industries and automation (5G-ACIA), and the critical communication association (TCCA) proffer regular inputs to the 3GPP. The main objective is to break historic silos between industrial and wireless communities in designing the beyond 5G networks according to the industrial needs~\cite{ho2019next}.
Many industries have opted for 5G for OT connectivity to achieve secure and safe manufacturing, productivity and efficiency~\cite{IET2019}.
From~\cite{Qualcomm2019,IET2019}, it is evident that private 5G networks are provisionally designed for industrial use cases, that can provide industries standalone dedicated resources and services rather than conventional mobile networks. It creates an opportunity for employing DTs at the 5G radio access network (RAN) layer with dedicated computational resources available and closer to the factory floor to learn, predict and make the decisions locally while communicating with the CDT in the cloud (Fig.~\ref{maindiagram}, \textit{Cloud Layer}). Moreover, the vital advantage for enterprise users in this new approach is designing the private and reliable mobile network according to their needs, which can satisfy the broad-scale coverage, stringent latency and reliability requirements, and security of industrial communication~\cite{Qualcomm2019}.
\begin{figure*}[!t]
\begin{center}
  \includegraphics[width=0.68\linewidth]{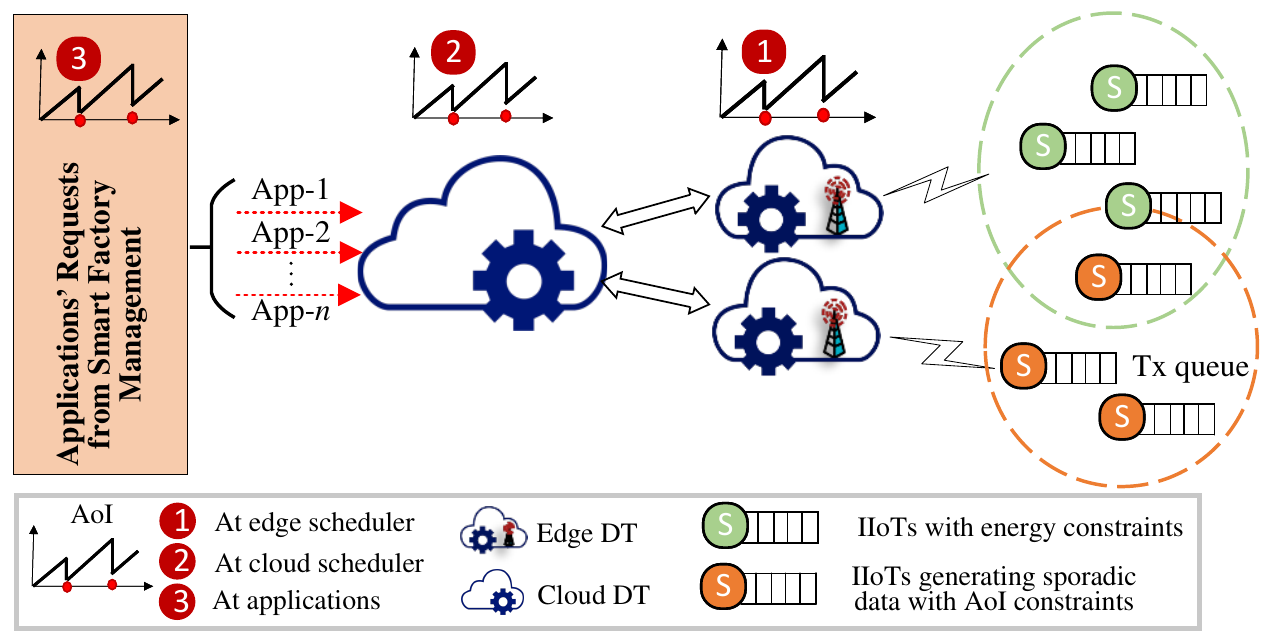}
  \caption{Scheduling of devices and resource allocation using CDT and EDT at different network levels depending on the energy consumption and AoI requirement set by the various applications' requests (factory management layer).}
  \label{fig:AoI}    
  \vspace{-15pt}
\end{center}
\end{figure*}
% The EDTs can also facilitate the B5G network performance by optimizing the resources for a single service, e.g., high bandwidth allocation using millimeter-Wave (mmWave) digital beamforming or employing multi-input multi-output (MIMO) antennas to provide eMBB service to camera operated remote control operations, or dynamically optimize network slicing to simultaneous support traffic of all three 5G network services, i.e., eMBB, URLLC, and mMTC~\cite{B5G2}. 
The EDTs can also facilitate the B5G network performance by optimizing the resources for: 1) a single service, e.g., high bandwidth allocation at millimeter-Wave (mmWave) bands or employing digital beamforming at multi-input multi-output (MIMO) antennas to provide eMBB service to camera operated remote control operations, or 2) dynamically optimize network slicing to simultaneous support traffic of all three 5G network services, i.e., eMBB, URLLC, and mMTC~\cite{B5G2}.
Meeting these requirements by twin-enabled 5G networks is fundamental in realizing the new era of mission-critical applications.

EDTs can also efficiently distribute standard time, e.g., universal time coordinated (UTC) information, to all the factory machines for synchronizing M2M communication. Three types of M2M sync are to be achieved among factory machines, as shown in Fig.~\ref{maindiagram}. Moreover, the propagation delays of signals from gNB to devices can be adjusted using a timing advance (TA) mechanism to estimate over-the-air propagation delays~\cite{AoI6}. The reduction in latency and increase in the reliability of these approaches increases the successful dissemination of new and periodic data from the machines to the nearest edge server running the twin-enabled 5G networks~\cite{MEC3}.
\subsubsection{Multi-access Edge Computing}
\label{sec:MEC}
The significance of edge computing architecture from the DT's perspective is explored and discussed in Sec~\ref{sec:E-DT}.
Meanwhile, the other emerging trends in B5G networks is "multi-access edge computing (MEC)" which is a part of edge computing techniques in which 5G gNBs are integrated with the computation and storage resources~\cite{MEC1}. 
This approach benefits in decreasing the application latency, traffic congestion at local mobile networks, and improving the end-user's quality of experience (QoE) and QoS by moving the cloud computing capabilities to the edge of 5G RAN.

Fig.~\ref{maindiagram} (\textit{Edge Layer}) shows the MEC concept for DT-enabled smart industry in which the generated factory data is offloaded, stored, and computed at MEC of the 5G RAN layer.
Combining MEC at 5G RAN with ML and AI model (shared by the cloud using federated learning approach) and local analytics greatly benefit in developing the new agile class of EDTs for the smart industries, which learns and renders the simulation of the entire local manufacturing facility through the inflow of large volume of data at 5G gNB~\cite{MEC3}. 
As discussed in the previous Section~\ref{sec:5GNWservices}, these new classes of EDTs at 5G RAN, trained on real-time data of the local facility, can provide better network services and fulfill various constrained requirements of use cases mentioned in Table~\ref{table:I}.

\subsubsection{Green Communication}
\label{sec:GreenCommunication}
Smart factory is tied to denser and wide-scale monitoring and control of sensors and actuators through low-complexity IIoT devices, often deployed in harsh and inaccessible locations without grid power~\cite{zeb2020impact,zeb2021impact}. In such scenarios, communication of battery-operated devices needs to be carefully optimized since communications are typically the most energy-draining operation. Additionally, offloading of computations workload from end-users (machines) to edge layers at 5G RAN or dedicated standalone edge server can save industrial sensors' power consumption, extending their battery life~\cite{Gr2}. However, solutions based on battery-operated devices suffer from various concerns such as network lifetime, and environment unfriendly and costly battery recycling and replacements, respectively. To overcome the battery-related challenges, different energy harvesting (EH), wireless power transfer, and backscattering-based wireless networks are being investigated~\cite{Gr4,Gr6,Gr7, Nazar_TGCN_BC}. The challenge remains on the refresh rate of EDTs and CTD, which, depending on the application, affects the querying rate of devices and might mismatch with their energy renewal rate.
%%%%%%%%%%%%%%%%%%%%%%%%%%%%%%%
\subsubsection{Age of Information}
\label{AoI}
Numerous industrial applications rely on the updated data collection over time, while the data must possess the property of having new and fresh information, i.e., the minimum AoI~\cite{AoI1,abbas2021age}. Table~\ref{table:I} shows the update time of generated periodic traffic in various industrial use cases.
%AoI is the performance metric associated with data, which indicates its freshness,i.e., data age since it was created during factory operations. 
%AoI of the machines data at the factory floor form the performance bottleneck for the smart industry applications as machines in factory processes generate the data that carry essential information, which has a high value at the upper layers for further taking appropriate action~\cite{AoI2}. 
% AoI of the collected data forms the performance constraint in critical decision-making process as the data carries high value essential information regarding machine processes~\cite{AoI2}.
% Table~\ref{table:I} shows the update time of generated periodic traffic inflow from machine devices in various industrial use cases. Periodic and sporadic data updates are cached at the edge servers accessible by the upper layer applications, resulting in minimum AoI~\cite{AoI3}. However, because of EH constraints and scarce radio resources under the vision of green communication, not being to entertain and reply to all sensing (application) requests from the factory terminal layer (illustrated in Fig.~\ref{fig:AoI}) leads to deterioration in the AoI metric of sensed data.
AoI of the collected machine data forms the important performance metric for critical decision-making processes as the data value reduces with the elapsed time~\cite{AoI2}. 
Hence, minimum AoI is desired for reliable and agile decisions. For this purpose, one strategy is to cache the (periodic and sporadic) data updates from IIoTs at the edge servers, which are readily accessible to the upper layer applications, resulting in minimum AoI~\cite{AoI3}. This problem introduces the tradeoff for cloud and edge-based strategies in providing data updates with either minimum data AoI, achieved by frequently polling each machine and sensing device, or with aged information from cached data at the server.  
As discussed in Section~\ref{sec:E-DT}, EDTs and CDTs have access to the incoming periodic data traffic stored in the cache with different AoI. The authors in \cite{AoI4} proposed the AoI-aware scheduling policy together with learning at EDT and CDT, which can dynamically address the tradeoff between minimum AoI of data and cached data from field devices in a large wireless network. The next-best approach to disseminating the new factory data information in multi-hop wireless sensors networks is cooperative communication~\cite{AoI5}. This approach dramatically improves information packets' reliability by reaching the sensor network's gateway (i.e., 5G gNB) in any direct communication failure.

On the other hand, because of EH constraints (or green communications) of IIoTs and scarce radio resources, not being able to entertain and reply to all sensing (application) requests from the factory terminal layer (illustrated in Fig.~\ref{fig:AoI}) leads to deterioration in the AoI metric of sensed data. Therefore, it is crucial to develop intelligent network slicing schemes with multi-objective resource allocation criteria. Each objective must capture the requirements and requests of industrial applications realistically using critical metrics as service rate, scheduling and isolation, information freshness, and energy efficiency. In this direction, in \cite{abedin2021elastic}, using distributed game theory and machine learning, the authors developed an elastic network slice policy to satisfy time-varying resource allocation demands for three different industrial traffic classes. In the slice configuration policy, the authors mainly aimed to balance AoI and energy efficiency while maximizing the service rate.
For this approach to work, M2M timing synchronization (M2M sync) between the multiple collaborating machines (shown in Fig.~\ref{maindiagram}, \textit{Physical Resource Layer}) is crucial~\cite{AoI6}. 
%For this approach to work, time synchronization between the multiple collaborating machines is of great necessity~\cite{AoI6}. 
%%%%%%%%%%%%%%%%%%%%%%%%%%%%%

% Past research indicates that ML techniques have widespread adoption to facilitate automation in industrial applications.
%%%%%%%%%%%%%%%%%%%%%
\section{Lessons Learned, Challenges, and Future Directions}
\label{finalsection}
In this section, we discuss various important observations, recommendations, and open future research challenges/problems associated with the practical utilization of industrial DT in conjunction with the emerging communication and computation technologies.

\subsection{Lessons Learned and Challenges}
\label{sec:LessonsLearned}
Based on the systematic review presented in the previous sections, the key practical lessons and recommendations are as follows.
\begin{itemize}
    \item It is clear from the summarized review (c.f.~Sec.~\ref{sec:IDTsec}) that industrial DT has been mostly explored for the factory processes associated with robotic production, prognosis, and devices health management applications. The inclusion of DT technology in the applications, as mentioned earlier, brings significant gains over traditional optimized methods (physical modeling, geometrical modeling) due to the further incorporation of command and management/behavior modeling aspects in living softwarized replica built upon the input incurred IIoT data. Implementing and simulating these DT-driven softwarized models utilizes the fused (physical and virtual) data, past historical data, real-time data, and simulation data, resulting in an accurate depiction of practical situations. It enables back and forth digital hyperconnectivity support between factory floor machines (physical entities) and softwarized replicas, leading to an agile decision-making process. Nevertheless, contemporary research on DT focuses more on a single machine and/or equipment. However, this limits the scope of industrial DT applicability to the entire manufacturing floor covered with multiple collaborating robots and operational machines, which requires exploring trending ICT techniques to facilitate the industrial DT's build-up.    
    Moreover, there is no consensus on the general design framework for industrial DT; therefore, a unified research and development effort is needed towards DT implementation.
    % Moreover, there is no consensus on the general design framework for industrial DT, due to which the unified development and research lag towards the common challenging goal of DT implementation. We must explore and address these limitations to bridge the lagging research and development towards industrial DT.
    \item The integration of industrial DT technology with trending cloud/edge-based data computing methods and enhanced core network connectivity through software-defined networking (SDN) and network function virtualization (NFV) technologies are paving the way to move cloud DTs (CTDs) closer to the factory manufacturing floor with edge DTs (EDTs). This move will deliver exciting new enhanced security and privacy features, high reliability and low latency, and an agile-decision-making processes. Moreover, providing cloud computing capabilities at the edge layer can provide the deployment infrastructure for cloud-based microservices, which is easily accessible by EDTs to enhance its operational capabilities in processing IIoT data from heterogeneous sources, performance monitoring, and optimizing the factory processes. However, the computation capabilities and networked connectivity (wireless and wired) at the edge and cloud layer to support the CDT and EDT-driven factory operation is far challenging because of: 1) insufficient computation resources, 2) non-optimized network architecture to support high data traffic flows, 3) complexities in software and hardware configurations of networking infrastructure, 4) DT-aware network communication protocol, and 5) the geographical distribution of clouds.
    \item Data fusion, acquisition, and mining will play an essential part in giving true meaning to CDT and EDT function realization. Altogether, these techniques will effectively link the cyber and physical space of the manufacturing floor by simultaneously processing and fusing the multiple features of acquired time-series machine data from multiple heterogeneous sources (physical space) with past data records, behavior, and simulation data (cyber-space). However, data fusion and mining at such a massive scale for CDT and EDT incite the availability and implementation challenges of robust, computationally efficient, and resilient algorithms that can accurately model and fuse the DT data. 
    \begin{figure*}[!t]
    \centering
    \includegraphics[width=0.85\linewidth]{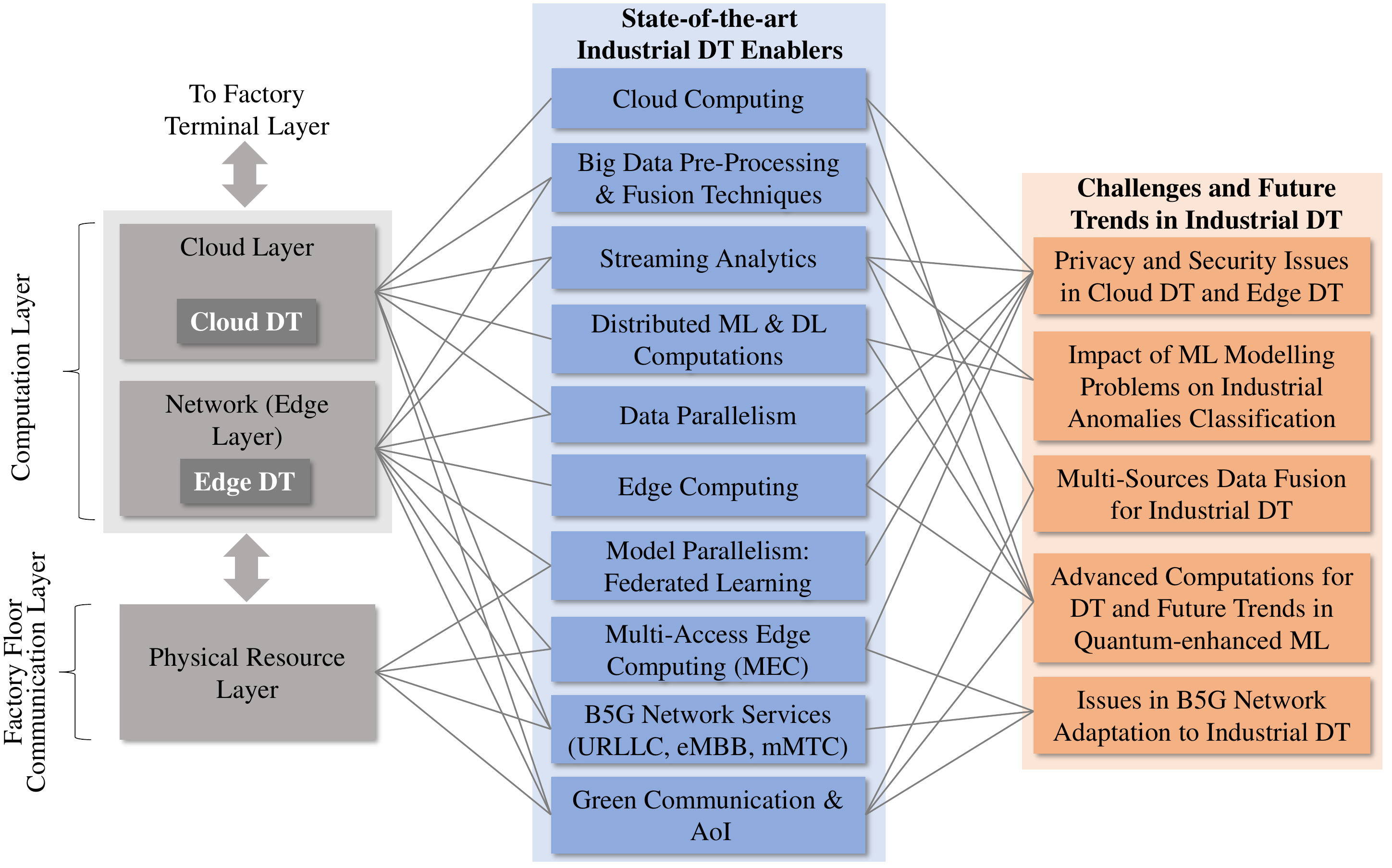}
    \caption{An overview and mapping between factory communication stack, state-of-the-art enablers, and challenges and future trends for Industrial DT.}
    \label{fig:Overview}
\end{figure*}
    \item The emergence of novel ML and DL frameworks with a de-facto hybrid cloud-edge-native computing architecture will play an integral part in stimulating the numerous functional aspects (i.e., prognosis, simulation) of CDTs and EDTs. Especially the use of streaming data analytics and distributed federated learning-based computation techniques will significantly improve the security, privacy, low latency, and reliability aspects at CDT and EDT. It will undoubtedly enhance the data computations of the incurred data, accurate prognosis, and agile decisions in DT-driven industrial processes. However, many challenges exist for the complete adoption of ML and DL frameworks for industrial DT applications: 1) large high-quality labeled data for training ML/DL algorithms, which increases the importance of preprocessing raw data and data fusion techniques, 2) current networking infrastructure lags the computational support for AI-based service deployment and solutions, 3) malicious exchanges of learning model updates during the federated learning process can affect the integrity of DT model, and 4) local and global aggregation of the learned models at the EDT and CDT varies from application to application.
    \item From the discussion in Section.~\ref{5GRole} and Fig.~\ref{5G}, it is clear that B5G networks will play an essential role in providing wireless connectivity services and URLLC, eMBB, and mMTC-service support between industrial DT (CDT and EDT) and physical space (manufacturing floor). Moreover, the innovative edge intelligence vision of 6G is all about integrating state-of-the-art AI services and hybrid edge-native computing models to provide on-demand services to various applications, which will undoubtedly extend the service-based architecture of B5G to new heights. This will prove beneficial in terms of supporting the objectives of Industrial DTs, e.g., softwarization of industrial processes, support for data computations near the edge layer (MECs), optimized resources for wireless data transfer in a harsh indoor multipath-riched industrial environment, AR/VR support for visualization. However, there are critical challenges in developing standards and fully adopting the emerging 5G technologies and enablers (SDN/NFV, mmWave bands, MIMO, MEC, etc.) in B5G/6G cellular networks to build a universal B5G-based wireless ecosystem for industries as they are not technologically matured enough to provide industrial-grade connectivity.    
    \item In reaching the objectives of industrial DTs, the sensing, communication, and computing ecosystem will require adopting a holistic approach towards energy-efficient, green infrastructure. As dense sensing leads towards massive battery-operative and energy harvesting (EH) devices, new hardware, communication infrastructure, and algorithmic approaches are needed. Meanwhile, due to the critical nature of industrial DTs, the QoS must be objectively captured and maintained for industrial DTs. Therefore, radio access, core network, and computing resources have to be jointly optimized while balancing the trade-off between energy efficiency and QoS. At radio access, energy-efficient schemes include (but are not limited to) link adaptation and topology, radio resource, and network management techniques. At the core network, energy efficiency can be enhanced by dynamic resource activation and virtualization. Meanwhile, the green computing solutions range from deployment optimization (e.g., at cloud or edge) to virtualization of computing resources. Although cloud computing lowers the energy cost, distributed edge computing can meet industrial QoS targets and facilitates UEs' energy efficiency by computation offloading. Importantly, AoI, which objectively captures the value of information for industrial control and monitoring applications, should be at the center place for various tasks, such as communication and computing resource allocation, device polling and scheduling, and energy sources.
\end{itemize}

\subsection{Future Research Directions}
\label{sec:future}
In the previous sections, we identified the critical challenges, i.e., security, privacy, big data computations, data updates, and communication constraints for complete utilization of DT adoption in industrial scenarios. Moreover, we discussed the role and requirements of emerging technologies with respect to catering and facilitating the seamless integration of DT at each layer of the factory communication stack in Fig.~\ref{maindiagram}. The provision of emerging technologies to meet the demands of performance metrics in the industrial DT landscape creates a plethora of research directions, which must be explored in the future. 
In Fig.~\ref{fig:Overview}, we give an overview and mapping between factory communication stack, discussed state-of-the-art features and enablers, and challenges/future trends for Industrial DT. This section highlights the foreseeable open research problems in DT for the smart industry at edge and cloud layers, 5G adaptation, data-related issues, advanced computation, and classification problems.
\subsubsection{Privacy and Security Issues; Blockchain Technology}
\label{sec:PandS}

Integration of critical technologies on factory floor like IIoT and CPS-based factory machines and 5G networks have made it relatively easy to sinew a range of industrial devices communicate wirelessly and enable them to share data ubiquitously even from distant locations. 
The computation servers at cloud and edge layers where DT resides are inundated with high-value and larger-volume  periodic and sporadic data from complex industrial processes.
Therefore, there is a strong need to defend against the security and privacy issues arising from the perpetrators, unauthorized machine access, remote attacks, and rival intruders at each factory communication stack layer. Protecting these machine communication carrying high-value data is challenging.

Blockchain technology can be an effective solution for DT in smart industries to maintain industrial data integrity in virtual cyberspace, bringing future research opportunities. A chain of blocks, rechristened as blockchain, is a growing list of records called a block linked with all other blocks using cryptography. A motif of these sophisticated blocks is intrinsically disciplined to resist any possible data modification, or any tampering to industrial data at any point in these blocks is inherently inviolable. Ledgers of the blockchains are designed altogether ingeniously and holistically, that if any addition is once made, it can never be edited or deleted and the hash of the block acts as DNA. Thus, the provided feed to EDTs and CDTs is secure and intact.

\subsubsection{Major Challenges in Adaptation of 5G for Industrial DT}
\label{sec:5GIssues}
% \textcolor{red}{
% Integrating 5G and DT for smart manufacturing indeed opens up a new avenue of opportunities. However, several challenges in the adaptation of B5G networks hinder DT's development process at the edge and cloud layer. Some of these challenges are:
% }
Integrating B5G networks and DT technology for smart industries opens up a new avenue of futuristic opportunities. However, several research challenges exist in adapting B5G networks to meet the time-critical applications' communication constraints that hinder the DT's development process at the edge and cloud layer. Some of these highlighted open research challenges are:
\begin{enumerate}%[leftmargin=*]
    \item Software-defined networking (SDN) and network function virtualization (NFV) are the key enablers in realizing the 5G network slicing architecture to fulfill the applications' diverse needs and requirements. However, there is a need to explore SDN and NFV functionality for 5G network slicing techniques to support the DT requirements.
    \item The interplay of mmWave and MIMO technology along with the availability of a dedicated licensed spectrum for industries is needed to provide the demanding URLLC and eMBB services. 
    \item Currently, most of the literature primarily focuses on addressing the network side concerns of 5G in a smart factory. However, there is a limited progress on the 5G-enabled industrial devices, e.g., 5G connectivity chipset modules for industrial devices, and their seamless compatibility with other industrial communication technologies such as Ethernet and field buses.
\end{enumerate}

\subsubsection{Modeling Problems in Anomalies Classification}
\label{sec:Issues}
The prognosis of anomalies or fault detection in machines involved with complex manufacturing processes is typically classified as logistic regression problems that predict the class of malignant event occurrence in machines. However, the frequency class of faulty events is less in a real-time factory environment, which prevents the formation of a logistic classifier that can accurately predict the faulty events from provided machine data. The classifier will be more biased towards predicting the majority class of benign events, i.e., machines' regular operation, with greater accuracy while the less frequent class corresponding to critical fault or anomaly in machines is misclassified or ignored. This recurrent problem of misclassifying anomalies detection brings the wrong insights to DT for automated decision-making.
%, especially in complex manufacturing processes of smart industries use cases with critical URLLC and eMBB communication requirements. 
To address the class imbalance in training data set of machines data, preprocessing methods such as class undersampling and oversampling techniques, or embedded modifications in the model of the ML framework can be explored from the DT's perspective.

\subsubsection{Challenges in Multi-source Data Fusion for DT}
\label{sec:Fusion}
Data fusion techniques will play an integral part in facilitating EDT and CDT to handle and fuse the multi-heterogeneous big industrial data so that the fused data gives an optimized insight to the factory twin. Besides data fusion, aggregated knowledge of the machines from past historical data accumulated over the years from complex industrial processes and technical knowledge from experts is also instrumental in building an accurate prognosis model at EDT and CDT. 
There is a limited research on the integration of all these factors, meant to expedite DT building process in smart factories.
%There has been limited work done regarding addressing this essential issue in integrating all these factors to facilitate the DT building process for the smart factory.  

\subsubsection{Quantum-enhanced Machine Learning (QML)}
\label{sec:QC}
Recent advances in the computation field has led to new QML architecture. QML has emerged from merging two interdisciplinary research areas: \textit{ML and quantum physics}. It deals with executing state of the art ML algorithms on classical data using the quantum computer. QML can increase the computation power on big data of smart factories by intelligently analyzing data in the realm of quantum states. The integration of QML algorithms at cloud or edge will undoubtedly give fast and accurate updates to CDT and EDT. Moreover, a hybrid processing mechanism can be explored from the perspective of DT usage in which complicated subroutines of computation processes are assigned to the quantum devices for faster execution, while at the same time, the rest is fed to the conventional computational server machines.

%%%%%%%%%%%%%%%%%%%%%%%%%%%%
\section{Conclusion}
\label{sec:conclusion}
This paper reviewed DT usage in the intelligent industry scenario and presented enabling computing and communication techniques in NextG wireless networks and computational intelligence paradigm. The ever-evolving world of industrial communication is becoming dynamic thanks to the introduction of numerous emerging technologies. This incurs the enabling requirements for the successful implementation of intelligent processes in the factories using industrial DT. Often, the misuse of the term CPS and DT is expected in the context of Industry 4.0. In this paper, we initially provided a systematic review of DT usage specifically for the smart industries and described its significance in terms of value and impact in revolutionizing the concept of intelligent services in Industry 4.0. Afterward, we identified emerging technologies' critical role and requirements, i.e., cloud and edge computing, ML and data analytic techniques, green communication and AoI, and Beyond-5G networks, in the DTs for smart industries. We also discussed the various advances and concepts within these technologies that intelligent industries can exploit for realizing the new class of DTs.
% Besides, the smart services that characterize the performance of intelligent industries with DT, it still bears the challenges stemming from the requirements possessed by emerging technologies in terms of security, privacy, low latency, high reliability, computations, and freshness of data from devices.
Besides the intelligent services enabled by DT in the industries, it still bears the challenges stemming from the critical requirements possessed by emerging technologies, i.e., privacy and security, major challenges in adaptation of 5G-and-beyond networks, anomalies classification, multi-source data fusion, and enhanced machine learning.

%%%%%%%%%%%%%%%%%%%%%%%%%%%%
\ifCLASSOPTIONcaptionsoff
  \newpage
\fi
%%%%%%%%%%%%%%%%%%%%%%%%
%\bibliographystyle{IEEEtran}
%\bibliography{Biblio}
% Generated by IEEEtran.bst, version: 1.14 (2015/08/26)

%%%%%%%%%%%%%%%%%%%%%%%%
\vspace{-0.5cm}
%%%%%%%%%%%%%%%%%%%%%%%%
%\begin{IEEEbiography}[{\includegraphics[width=1in,height=1.25in,clip,keepaspectratio]{images/shahzeb (1).jpg}}]{Shah Zeb}(Student Member, IEEE) received his B.E. degree
\begin{IEEEbiography}[{\includegraphics[width=1in,height=1.25in,clip,keepaspectratio]{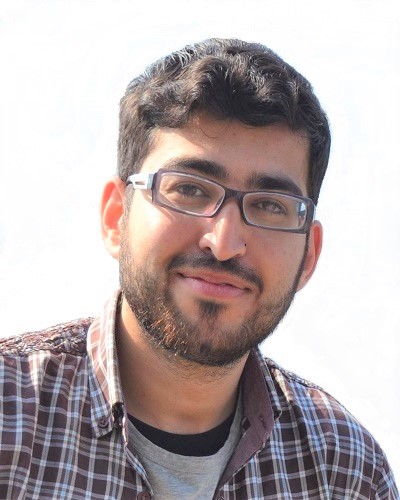}}]{Shah Zeb} received his B.E. degree in Electrical Engineering from the University of Engineering and Technology at the Peshawar Campus, Pakistan, in 2016. He received his M.S. degree in Electrical Engineering from National University of Sciences and Technology, Pakistan, in 2019. He is currently pursuing the Ph.D. degree with the National University of Sciences and Technology (NUST), Pakistan. Also, he is currently working as a Research Associate with the Information
Processing and Transmission (IPT) Lab, School of Electrical Engineering and Computer Science (SEECS), NUST, which focuses on various aspects of wireless communications. His current research interests include ultra-reliable and low-latency communication for Industrial IoT, Digital Twin, millimeter-Wave communication, backscatter communications, and NOMA. 
\end{IEEEbiography}

% % %%%%%%%%%%%%%%%%%%%%%%%%

\begin{IEEEbiography}[{\includegraphics[width=1in,height=1.25in,clip,keepaspectratio]{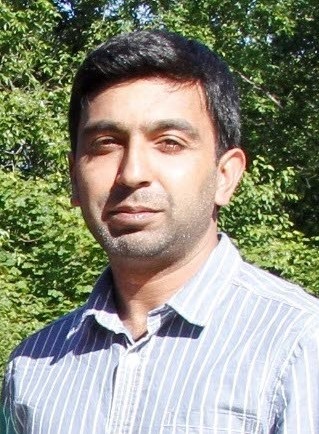}}]{Aamir Mahmood}
received the B.E. degree in electrical engineering from the National University of Sciences and Technology, Islamabad, Pakistan, in 2002, and the M.Sc. and D.Sc. degrees in communications engineering from the Aalto University School of Electrical Engineering, Espoo, Finland, in 2008 and 2014, respectively. He worked as a Research Intern with Nokia Research Center, Helsinki, Finland, in 2014, a Visiting Researcher with Aalto University from 2015 to 2016, and a Postdoctoral Researcher with Mid Sweden University, Sundsvall, Sweden, from 2016 to 2018, where he has been an Assistant Professor with the Department of Information Systems and Technology, since 2019. His research interests include low-power local/wide-area networks, energy-delay-aware radio resource allocation, and RF interference/coexistence management.
\end{IEEEbiography}
% % %%%%%%%%%%%%%%%%%%%%%%%%
% % %%%%%%%%%%%%%%%%%%%%%%%%

\begin{IEEEbiography}[{\includegraphics[width=1in,height=1.25in,clip,keepaspectratio]{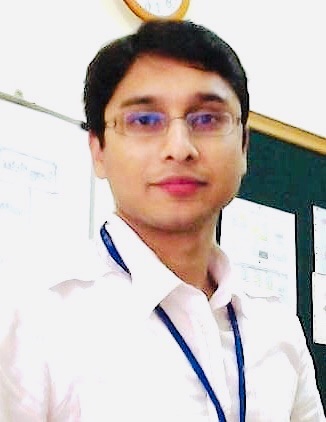}}]{Syed Ali Hassan}
 received the B.E. degree in electrical engineering from the National University of Sciences \& Technology
(NUST), Islamabad, Pakistan, in 2004, the M.S. degree in mathematics from Georgia Tech in 2011, and the M.S. degree in electrical engineering from the University of Stuttgart, Germany, in 2007, and the Ph.D. degree in electrical engineering from the Georgia Institute of Technology, Atlanta, USA, in 2011. 
His research interests include signal processing for communications with a focus on cooperative communications for wireless networks, stochastic modeling, estimation and detection theory, and smart grid communications. He is currently working as an Associate Professor with the School of Electrical Engineering  and Computer Science (SEECS), NUST, where he is the Director of the Information Processing and Transmission (IPT) Lab, which focuses on various aspects of theoretical communications. 
He was a Visiting Professor with Georgia Tech in Fall 2017. He also held industry positions with Cisco Systems Inc., CA, USA, and with Center for Advanced Research in Engineering,
Islamabad.
\end{IEEEbiography}
% % %%%%%%%%%%%%%%%%%%%%%%%%

\begin{IEEEbiography}[{\includegraphics[width=1in,height=1.25in,clip,keepaspectratio]{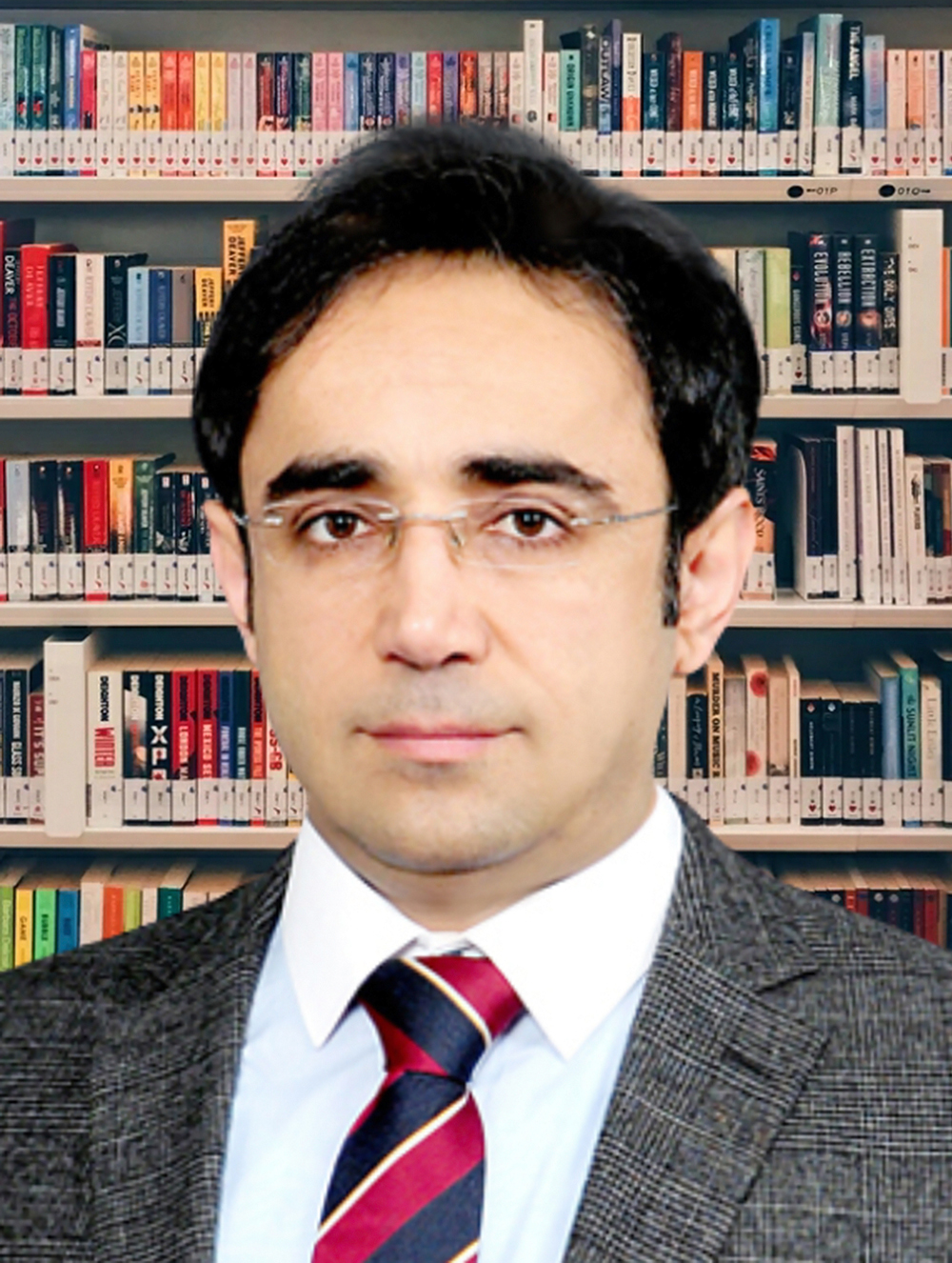}}]{MD. Jalil Piran}
is an Assistant Professor with the Department of Computer Science and Engineering, Sejong University, Seoul South Korea. Jalil Piran completed his PhD in Electronics
Engineering from Kyung Hee University, South Korea, in 2016. Subsequently, he continued his work as a Postdoctoral Research fellow in the field of ”Resource Management” and ”Quality of Experience” in ”5G and beyond” and ”Internet of Things”
in the Networking Lab, Kyung Hee University. Dr. Jalil Piran published substantial number of technical papers in well-known international journals and conferences in research fields
of ”Wireless Communications and Networking,” ”Internet of Things (IoT),” ”Multimedia Communication,” ”Applied Machine Learning,” ”Security,” and ”Smart Grid”. He received ”IAAM Scientist Medal of the year 2017 for notable and outstanding research in the field of New Age Technology \& Innovation,” in Stockholm, Sweden. Moreover, he has been recognized as the ”Outstanding Emerging Researcher” by the Iranian Ministry of Science, Technology, and Research in 2017. In addition, his PhD dissertation has been selected as the ”Dissertation of the Year 2016” by the Iranian Academic
Center for Education, Culture, and Research in the field of Electrical and Communications Engineering. In the worldwide communities, Dr. Jalil Piran is an active member of Institute of Electrical and Electronics Engineering (IEEE) since 2010, an active delegate from South Korea in Moving Picture Experts Group (MPEG) since 2013, and an active member of International
Association of Advanced Materials (IAAM) since 2017.
\end{IEEEbiography}

% % %%%%%%%%%%%%%%%%%%%%%%%%%
\begin{IEEEbiography}[{\includegraphics[width=1in,height=1.25in,clip,keepaspectratio]{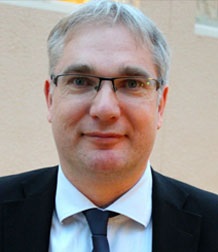}}]{Mikael Gidlund}
 received the Licentiate of Engineering degree in radio communication systems from the KTH Royal Institute of Technology, Stockholm, Sweden, in 2004, and the Ph.D. degree in electrical engineering from Mid Sweden University, Sundsvall, Sweden, in 2005. From 2008 to 2015, he was a Senior Principal Scientist and a Global Research Area Coordinator of Wireless Technologies with ABB Corporate Research, Västerås, Sweden. From 2007 to 2008, he was a Project Manager and a Senior Specialist with Nera Networks AS, Bergen, Norway. From 2006 to 2007, he was a Research Engineer and a Project Manager with Acreo AB, Hudiksvall, Sweden. Since 2015, he has been a Professor of Computer Engineering at Mid Sweden University. He holds more than 20 patents (granted and pending applications) in the area of wireless communication. His current research interests include wireless communication and networks, wireless sensor networks, access protocols, and security. 

Dr. Gidlund is an Associate Editor of the IEEE TRANSACTIONS ON INDUSTRIAL INFORMATICS.

\end{IEEEbiography}
%%%%%%%%%%%%%%%%%%%%%%%%%
\begin{IEEEbiography}[{\includegraphics[width=1.25in,height=1.25in,clip,keepaspectratio]{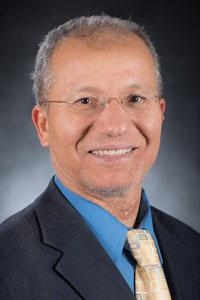}}]{Mohsen Guizani}
received the B.S. (Hons.) and M.S. degrees in electrical engineering and the M.S. and Ph.D. degrees in computer engineering from Syracuse University, Syracuse, NY, USA, in 1984, 1986, 1987, and 1990, respectively. He is currently a Professor with the Computer Science and Engineering Department, Qatar University, Qatar. Previously, he has served in different academic and administrative positions at the University of Idaho, Western Michigan University, the University of West Florida, the University of Missouri–Kansas City, the University of Colorado at Boulder, and Syracuse University. He is the author of nine books and more than 600 publications in refereed journals and conferences. His research interests include wireless communications and mobile computing, computer networks, mobile cloud computing, security, and smart grid. He is a Senior Member of ACM. He has served as a member, the Chair, and the General Chair for a number of international conferences. Throughout his career, he received three Teaching Awards and four Research Awards. He was a recipient of the 2017 IEEE Communications Society Wireless Technical Committee (WTC) Recognition Award, the 2018 AdHoc Technical Committee Recognition Award for his contribution to outstanding research in wireless communications and Ad-Hoc Sensor Networks, and the 2019 IEEE Communications and Information Security Technical Recognition (CISTC) Award for outstanding contributions to the technological advancement of security. He was the Chair of the IEEE Communications Society Wireless Technical Committee and the Chair of the TAOS Technical Committee. He is currently the Editor-in-Chief of IEEE Network Magazine, serves on the editorial boards for several international technical journals, and the Founder and the Editor-in-Chief for Wireless Communications and Mobile Computing journal (Wiley). He guest edited a number of special issues in IEEE journals and magazines. He has served as the IEEE Computer Society Distinguished Speaker and is currently the IEEE ComSoc Distinguished Lecturer.

\end{IEEEbiography}
%%%%%%%%%%%%%%%%%%%%%%%%%
\end{document}